\documentclass[useAMS,usenatbib]{mn2e}
\usepackage{graphicx}
 
\title[LFs in MUSYC-ECDFS]{Introducing the Photometric Maximum Likelihood Method: Galaxy
  Luminosity Functions at $z<1.2$ in MUSYC-ECDFS\footnotemark[1]}

\author[D. Christlein et al.]{Daniel Christlein \\ Max-Planck-Institut f\"ur Astrophysik, Karl-Schwarzschild-Str. 1, 85748 Garching, Germany \\
Yale University, Department of Astronomy, 260 Whitney Avenue, New Haven, CT 06511\\
Universidad de Chile, Departamento de Astronom\'ia, Casilla 36-D, Santiago de Chile
\newauthor Eric Gawiser\\ Rutgers University, Department of Physics and Astronomy, 136 Frelinghuysen Rd, Piscataway, NJ 08854, U.S.A.
\newauthor Danilo Marchesini \\ Yale University, Department of Astronomy, 260 Whitney Avenue, New Haven, CT 06511
\newauthor Nelson Padilla \\ Pontificia Universidad Cat\'olica de Chile, Departamento de Astronom\'ia y Astrof\'isica, \\Vicu\~na Mackenna 4860, 7820436 Macul, Santiago de Chile}

\begin{document}
\maketitle
\begin{abstract}
We present a new maximum likelihood method for the calculation of galaxy luminosity functions
 from multi-band photometric surveys without spectroscopic data. The
 method evaluates the likelihood of a trial luminosity function by
 directly comparing the predicted distribution of fluxes in a
 multi-dimensional photometric space to the observations, and thus
 does not require the intermediate step of calculating
 photometric redshifts. We apply this algorithm to $\sim 27,000$ galaxies
 with $m_{R}\leq25$ in the MUSYC-ECDFS field, with a focus on recovering the
 luminosity function of field galaxies at $z<1.2$. Our deepest LFs
 reach $M_{r}\approx-14$ and show that the field galaxy LF deviates
 from a Schechter function, exhibiting a steep upturn at intermediate
 magnitudes that is due to galaxies of late spectral types.
\end{abstract}


\footnotetext[1]{accepted for publication by MNRAS}

\section{Introduction}
\label{sec:intro}
The galaxy luminosity function (LF) has long been one of the
cornerstones of statistical observational cosmology. As the
distribution of galaxies as a function of luminosity is closely
related both to the halo mass function predicted by structure
formation models and to the astrophysical mechanisms governing star
formation, the LF provides a fundamental test for models of galaxy
formation and evolution. The low-redshift LF has also been extensively studied (see
\citet{binggeli} for a review), and some convergence has been reached
on its general shape at the bright end. In the famous parametrization
by \citet{schechter76}, a faint-end power-law slope of
$\alpha\approx-1.2$ is the consensus value. The characteristic
bright-end absolute magnitude $M^{*}$ is dependent on the magnitude
system and filter used, but the determination by \citet{blanton2003}
sets it at $M^{*}=-21.18\pm0.04 $ in the $r$-band at z=0.1 (adjusted for $H_0=71$).

Nevertheless, the LF remains a field of ongoing study as modern instrumentation allows us to survey increasingly fainter galaxies. One aspect that has received great attention is the evolution of the LF with redshift, as this provides a much more stringent test of galaxy evolution models than the single data point that the local LF offers us. Several groups have studied the LF to limits of $z\approx1$ \citep{wolf} or even higher, up to and beyond $z\approx3$ \citep{poli,giallongo,marchesini}, and have found evidence for evolution. Broadly speaking, early-type galaxies show a marked number density evolution with redshift, while late-type galaxies exhibit a luminosity evolution. 

But another frontier in our exploration of the LF lies at low
redshifts and very faint absolute magnitudes. Traditionally, the
deepest probes of the LF come from studies of galaxy
clusters. Providing a sample of hundreds of galaxies at the same
distances, clusters offer an opportunity to determine LFs simply by
counting galaxies and then applying corrections for sample
contamination by unrelated background galaxies. This method has been
criticized by \citet{valotto} and shown to potentially produce
artificially steepened faint-end slopes for optically-selected
clusters \citep{valotto2} due to the fact that optical selection
favors clusters with a higher-than-average density of background
galaxies. However, since it eliminates the need for spectroscopic
observations to determine the exact distance of each galaxy, this
method has nonetheless been widely applied and has probed the LF of
nearby clusters to spectacular depths \citep{trentham98,trentham02a,trentham02b}. Frequently,
LFs calculated in this way show an upturn in the LF, setting in
several magnitudes below the characteristic bright-end magnitude M$^{*}$, from a slope of $\alpha\approx-1.2$ to a much steeper slope of $\alpha\approx -1.5$. More recently, a similar upturn has also been claimed by \citet{popesso} based on applying the background subtraction technique to clusters in the SDSS.
However, for a sample of clusters at $z\approx 0.3$, this has also recently been debated by \citet{harsono}, who are also working with statistical background subtraction, but find no evidence for an upturn in the cluster LF. Furthermore, \citet{rinesgeller}, working with a spectroscopic sample, find that the LFs of the Virgo Cluster and Abell 2199 can be represented by single Schechter functions consistent with the canonical values of $\alpha$ in the range of $-1.1$ to $-1.3$, and attribute the discrepancy from other studies to the use of statistical background subtraction without spectroscopic membership confirmation in the earlier work.
In principle, an upturn is not implausible. It is
known that galaxies with different morphologies or spectral types have
very different Schechter functions; for example, the LF of late-type
galaxies is generally very steep, while that of early-type galaxies is
shallower. If both are extrapolated to faint magnitudes and then
co-added, a faint-end upturn can be reproduced qualitatively
\citep{wolf}, provided that the normalization of the late-type LF is still high enough in these dense environments in order for it to dominate the overall LF at some point. Some surveys of nearby groups and clusters have also reported a ``dip'' in
the LF at intermediate magnitudes \citep{flint,miles2006,mendesdeoliveira06,biviano96}, which could also be an indication
of a bimodality that would be revealed as the superposition of a
shallow and a very steep Schechter function in a sufficiently deep and
large survey. However, the fact that spectroscopic surveys of the
field or clusters have not yet unambiguously verified this result (because of their generally shallower depth) makes this observation controversial. 

 It would be extremely important to know if this upturn is real and if
 it is specific to cluster environments (and thus an indicator of
 environment-specific processes generating large numbers of dwarf
 galaxies) or rather a universal feature of the LF. Current
 Cold-Dark-Matter cosmologies surmise that galaxies are embedded in
 dark matter halos, and the low-mass end of the halo mass function is
 thought to be quite steep, with a power-law slope of $\alpha\approx
 -1.8$. This is in marked contrast with the LF of field galaxies in
 the local universe, which exhibits a fairly flat faint-end slope of
 $\alpha\approx-1.2$ \citep{blanton2001,2dflf}. However, the field LF is well-established from
 spectroscopic surveys only to intermediate magnitudes of
 $M_{r}\approx-17$. Also, studies of the local LF in the only
 environment where we can probe its extreme faint end directly, the
 Local Group and the nearest of galaxy groups beyond it, prefer a flat
 faint end slope all the way down to the extreme dwarf regime of
 $M_{r}\approx-10$ \citep{pritchett,flint} or suggest at most a
 moderately steep slope \citep{trentham05}; these studies, however, suffer from being able to access only a very small cosmic volume. Although new Local Group members continue to be found up to the present day \citep{simongeha}, it is unlikely that they will change the census substantially at any but the very faintest luminosities, substantially fainter than the limit at which the steep faint-end upturn in clusters has been claimed to appear. Mechanisms suppressing the formation of galaxies in such low-mass halos \citep{koposov} may explain the small number of observable dwarf galaxies at low luminosities, but it is unclear how such mechanisms would account for the presence of a large dwarf galaxy population in clusters.
 
 A tentative suggestion of an upturn in the faint end of the field LF was found by \citet{blanton2005}, who detected a steepening of the slope to $\alpha=-1.3$ at intermediate luminosities, but speculated that inclusion of low-surface brightness galaxies undetectable by the SDSS might modify this value to $\alpha\approx-1.5$ or even steeper. Apart from this work, the aforementioned studies of the LF in clusters provide the only suggestion of a steeper faint-end slope so far, but these are questionable due to their samples lacking spectroscopic confirmation and therefore being susceptible to background contamination.

The frontier in exploring the LF at higher redshift and to fainter
absolute magnitudes therefore lies in pushing the flux sensitivity of
samples towards fainter and more distant objects. However, at such low
fluxes, spectroscopic redshift determination remains extremely
time-consuming and has a high failure rate, precluding its application
to large samples. As an alternative approach, photometric redshifts
are increasingly being substituted for spectroscopic redshifts even in
the study of LFs \citep{benitez,bolzonella}. However, it is well-known
that photometric redshifts are not an equivalent substitute for
spectroscopic redshifts; their accuracy is typically of the order of
$\Delta z=0.05(1+z)$ even in the best cases, and frequently worse than
that, leading to the galaxy distribution being smeared out in
luminosity and distance. In addition, they suffer from
catastrophic failures; galaxies of very different spectral types and
at different redshifts can have similar colors, leading to large
errors in their recovered photometric redshift that do not obey
Gaussian errors. Even purely Gaussian errors in the photometry
therefore inevitably propagate into non-Gaussian errors in the
photometric redshift, and samples selected by photometric redshifts are
inevitably contaminated by galaxies from galaxy populations
outside these selection limits. In addition, the distribution of
these errors varies from galaxy to galaxy, depending on its spectral
properties, true redshift, and magnitude. For all these reasons, the
recovered photometric redshifts and, by extension, LFs, are always
dependent on (explicit or implicit) prior assumptions about the LFs of
all galaxy populations at all redshifts --- assumptions which, except in the case of
very careful analyses, are often neglected, or in conflict with the
recovered results.

Unfortunately, luminosity function algorithms have adapted only poorly
to this situation. The algorithms applied to the problem of recovering
LFs from photometric-only samples are based on the same methods used
twenty years ago for the recovery of LFs from samples with known spectroscopic
redshifts. In the simplest procedure, photometric redshifts are simply
substituted for spectroscopic redshifts, and, assuming that the errors
in the photometric redshifts are sufficiently understood, Monte Carlo
simulations can be used to estimate whether a serious bias has been
introduced into the recovered LF parameters. This approach can be
successful, especially at high redshift, where a given uncertainty
$\Delta z$ in the photometric redshift translates into only a
relatively minor uncertainty in the absolute magnitude. An example of
a successful application of this procedure to the Multiwavelength
Survey by Yale-Chile (``MUSYC'')
\citep{gawiser_musyc} (http://www.astro.yale.edu/MUSYC) is found in \citet{marchesini}. However, at low redshift, as has been demonstrated in the past \citep{chen}, the problems associated with the redshift uncertainties are generally larger: a redshift uncertainty $\Delta z\approx 0.1$, typical for photometric redshift determinations, translates into a significant uncertainty in absolute magnitude. Correspondingly high-quality multi-band photometric data is needed to constrain the redshifts well \citep{wolf}, but obtaining such extensive datasets is not always practical.

There have been suggestions in the literature for adjustments to the
LF calculation that take into account the photometric uncertainties
and the resulting uncertainties in the redshift. These have either
employed Monte Carlo methods \citep{bolzonella} or incorporated the
photo-z uncertainties analytically into the procedure used for
recovering the LF \citep{chen}. However, both methods are based on the
idea of evaluating a redshift probability distribution that is
centered on the best-fit recovered photometric redshift, and {\it not
  centered on the true redshift}. The net effect is that the galaxy
distribution is subjected to a further convolution with the redshift
error function, instead of a deconvolution. 

The idea behind the {\it Photometric Maximum Likelihood} method
(hereinafter: PML), which we introduce in this paper, is therefore to
abandon the concept of assigning individual luminosities and redshifts
to individual galaxies, since neither are known in non-spectroscopic
surveys. Instead, we constrain the LF by requiring it to reproduce the
distribution of galaxies in a parameter space consisting of quantities
that are known exactly: the observed fluxes in multiple filter
bands. We refer to this parameter space, wherein observer-frame fluxes
in multiple filter bands are used as the coordinates, as {\it
  photometric space}. This algorithm achieves a true deconvolution of
the observed galaxy distribution into its constituent galaxy
luminosity functions, and it allows for a self-consistent
solution by avoiding priors that the recovered photometric redshifts
and luminosity distributions in photometric surveys normally depend on.

In this paper, we present the first application of the PML to a data set that is an ideal test case for this algorithm, the MUSYC survey \citep{gawiser_musyc}. MUSYC is a deep square degree survey (divided into four separate fields), comprising some $\sim 200,000$ galaxies. Photometric data in a multitude of optical and NIR filter bands exist. For the present study, we use one of the four MUSYC fields, the extended Chandra Deep Field South (E-CDFS), observed in eight filter bands: $U$, $B$, $V$, $R$, $I$, $z$, $J$, and $K$. The large number of filter bands make this survey very suitable to photometric redshift techniques as well as to the PML. The limiting depth of $R>26$ mag means that we can probe the LF to extremely faint limits; the relatively wide area, compared to other pencil-beam surveys, guarantees that even the bright end will still be fairly well-sampled at all but the smallest redshifts. With a sample cut at $R=25$, the PML permits us to constrain the field galaxy LF to a limit as faint as $M_{r}=-14$.

This paper is organized as follows: In \S \ref{sec_data}, we briefly
discuss the data set, the MUSYC survey. In \S \ref{sec_method}, we
present our LF algorithm the {\it Photometric Maximum Likelihood}
method and test it using a mock catalogue. In \S \ref{sec_results}, we present an application of our
method to MUSYC by calculating luminosity function parameters, split up by spectral type, from $z\approx 0.05$ to $z\approx 1.2$.

\section{Data}
\label{sec_data}

MUSYC is designed to provide a fair sample of the universe for the
study of the formation and evolution of galaxies and their central
black holes. The core of the survey is a deep imaging campaign in
optical and near-infrared passbands of four carefully selected
30$^{\prime}\times$30$^{\prime}$ fields. MUSYC is one of few surveys
to offer its combination of depth and total area, for additional
coverage at X-ray, UV-, mid-infrared, and far-infrared wavelengths,
and for providing the $UBVRIz^{\prime}JHK$ photometry needed for
high-quality photometric redshifts over a square degree of sky
(availability of NIR data varies from field to field). The primary
goal is to study the properties and interrelations of galaxies at a
single epoch corresponding to redshift $\sim3$, but the favorable
combination of depth, area, and passband coverage makes it suitable
for studies of the general galaxy population over a wide range of
redshifts as well as for Galactic astronomy \citep{martin}. 

Imaging data for three of the four MUSYC fields was obtained with
MOSAIC II on the 4 m Blanco Telescope at the Cerro Tololo Inter-American
Observatory (CTIO) in Chile. For the fourth field, the Extended Chandra Deep
Field South (E-CDFS), which we analyze in the present paper, our
$UBVRI$ imaging results from combining public images taken with the
ESO 2.2 m and WFI by the ESO Deep Public Survey and COMBO-17 teams 
\citep{erben,hildebrandt,arnouts,wolf2004}. Our $z^{\prime}$ imaging
was taken with the CTIO 4 m and MOSAIC II. Our $JK$ images of the E-CDFS were obtained with
the CTIO 4 m and ISPI.

Photometric coverage of the four MUSYC fields is not perfectly homogeneous. For
the present study, we have thus chosen to focus on one of the four
MUSYC fields, namely the E-CDFS,
which offers the deepest optical imaging; furthermore, in contrast to
the other MUSYC fields at the time of this writing, the optically
selected photometric catalogue with $UBVRIz$ photometry has been matched to NIR catalogues in
$J$ and $K$. The large number of spectroscopic redshifts available in
the literature also enable us to check the quality of our models by
estimating photometric redshifts. This field is centered on
$RA=03:32:29.0$, $Dec=-27:48:47$ (J2000). 

The data that we use for this analysis are drawn from a catalogue
selected by the combined $B$-, $V$-, and $R$-band flux, and
complemented by photometry in the $J$- and $K$-bands. The catalogues
of $J$- and $K$-band photometric data for the $BVR$-selected
detections have not been published, but separate, $K$-band-selected
catalogues of this field based on the same imaging material comprising
the NIR bands are already public and described in \citet{taylor}.

\begin{figure}
\includegraphics[width=8.4cm]{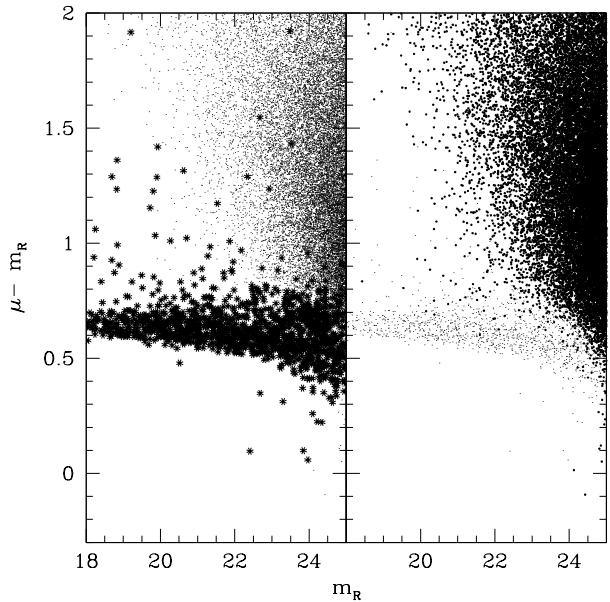}
\caption{Difference between average surface brightness (in mag
arcsec$^{-2}$) and magnitude $m_{R}$ versus $m_{R}$. Point sources
tend to populate a horizontal branch in this diagram, and extended
sources an extended cloud on the right side, but the two overlap for
faint magnitudes. In the left panel, objects with the star/galaxy
class $S/G>0.97$ are marked, in the right panel objects with
$S/G<=0.97$. This limit decomposes the two populations with minimal
overlap; we therefore accept objects with $S/G<=0.97$ as galaxies.
}
\label{fig_sg}
\end{figure}

The full catalog of the MUSYC E-CDFS contains 84410 objects. At the
depth of MUSYC, the detections consist primarily of galaxies, but the
catalog also contains point sources (stars and QSOs), which we remove
based on morphological criteria, using the Star/Galaxy classifier ($S/G$)
provided by the Source Extractor software \citep{sextractor}, which was used in the
construction of the detection catalog. To calibrate the threshold
value of $S/G$, we compare its distribution to the position of a
detection in the $m_{R}$-$\mu_{R}-m_{R}$ parameter space. In such a
parameter space, which is shown in Fig. \ref{fig_sg}, point sources
populate an approximately horizontal band. Experimentally, we find
that $S/G=0.97$ separates the two populations with minimal
overlap. This is demonstrated in Fig. \ref{fig_sg} by marking all
objects with $S/G>0.97$ with asterisks in the left panel, and all
other objects with dots in the right panel. This is a rather high
threshold, compared to values adopted in the literature in similar
efforts (i.e., $S/G=0.9$, \citep{capak}; $S/G=0.7$,
\citep{rowan-robinson}; $S/G=0.5$, \citep{grutzbauch}), and it is
possible that our S/G threshold fails to exclude scattered light from saturated stars or
close double star systems. To estimate the impact of the star/galaxy
threshold, we have therefore carried out a separate
analysis adopting a more conservative threshold of $S/G=0.8$ and will
compare its results with the default sample where appropriate in \S \ref{sec_discussion}.

For the calculation of the LF from a flux-limited sample, a thorough
understanding of the selection function is imperative. The optical
MUSYC source catalogs are detected by SExtractor in the
combined $B$-, $V$-, and $R$-band image, as described in \citet{gawiser_musyc}. The limiting magnitude for the E-CDFS field is
$BVR=27.1$ mag (in the AB magnitude system, which is used throughout
this paper). For the present paper, we
apply an additional cut in just a single filter band, the $R$-band, to
create our subsample. At $R\leq 25$ mag, corresponding to a flux limit of
$R=0.363$ $\mu$Jy, which is the limit adopted for our analysis, the
sample is complete in $BVR$. Objects with SExtractor warning flags 1,
2 (presence of a close neighbour object possibly affecting the
photometry; necessity of deblending) are left in
the catalog; objects with higher warning flags are removed without
statistical compensation. 

For our analysis, we use ``corrected aperture magnitudes''
\citep{gawiser_musyc}, which yield an estimate of the total flux, but
offer good S/N for compact sources. An exception are very extended
objects (with a half-light radius in the $BVR$-image of 1 arcsec or
more), for which we substitute the ``AUTO'' photometric measurement
provided by SExtractor. In addition, we have carried out modifications 
 to the published MUSYC photometry, as described in Appendix 
\ref{sec_photcorr}, to adjust the photometric zero points and ensure that
the photometric errors conform to a Gaussian
distribution. These corrections have significantly improved the
quality of the recovered photometric redshifts in comparison to a
subsample of galaxies with redshift estimates from the literature,
spanning a broad range of redshifts and SED types, and
have brought the uncertainty distribution in line with the assumptions
made in our algorithm.

\section{Method}
\label{sec_method}
\subsection{Principles of the Photometric Maximum Likelihood Approach}

The most common type of LF algorithms for magnitude-limited samples of field galaxies are maximum likelihood algorithms. These algorithms are based on the principle of calculating a predicted distribution of galaxies from a trial LF, comparing it to the observed distribution, determining the likelihood of the trial LF, and then modifying the latter until the likelihood has been maximized. Traditionally, the distributions that are being compared are the distributions over luminosity \citep{eep,sty}, sometimes over luminosity plus additional parameters that the LF could depend on, such as spectral type \citep{bromley,croton}, morphology \citep{binggeli}, or redshift \citep{wolf,poli,giallongo} itself.

In photometric surveys, neither spectral type, nor redshift, nor
luminosity are known exactly. Nonetheless, past attempts to calculate
the galaxy LF from photometric surveys have sought to reconcile the
data with the algorithm by simply substituting photometric redshifts
for spectroscopic redshifts. This can potentially cause several severe problems:
\begin{itemize}
\item {\it Random errors} in the photometric redshifts ``blur'' the recovered distances and absolute magnitudes. This introduces systematic biases in the recovered LF, as can be illustrated by a  {\it gedankenexperiment}: A typical LF is well-approximated by a power law at the faint end and has a well-defined exponential downturn at the bright end. If random errors exist in the galaxy redshift, the absolute magnitude estimates of all sample galaxies will be convolved with an error function, which ``blurs'' the shape of the LF and moves the bright-end ``knee'' to brighter luminosities. This can be minimized by operating at high redshift (where the impact of redshift uncertainties on the absolute magnitude is smaller) and, of course, with high-quality multi-band photometric data.
\item {\it Systematic errors and catastrophic failures} in the photometric redshifts may severely impact the recovered LF. For example, systematic offsets between the photometric redshifts and the true redshift (a common occurrence resulting from errors in the photometric calibration) will directly introduce shifts in the recovered absolute magnitudes. Catastrophic failures (i.e., photometric redshifts significantly and systematically different from the true redshifts) can have a devastating impact; for example, a small number of nearby, bright galaxies mistakenly fitted as high-redshift galaxies may easily dominate the bright end of the LF recovered at high redshift. It is important to note that such catastrophic failures can result simply from the propagation of Gaussian errors of the photometry into the photo-z determination. This means that modeling photometric redshift uncertainties as Gaussians is not legitimate even for perfectly-calibrated photometry.
\item {\it Sample contamination} is common. LF determinations are
  usually carried out for subsamples of galaxies in a confined
  redshift range and for selected SED types only. However, any
  subsample selected from photometric redshift data only will almost
  certainly be contaminated by galaxies at other redshifts and with a
  wide range of SED types, and incomplete with respect to galaxies
  that meet the selection criteria (see for example \citet{taylor}, which argues that photometric redshifts from the COMBO-17 survey \citep{wolf2004} suffer from contamination at $z>1$ due to the lack of NIR photometry). Aggravating this problem, the amount of contamination is, of course, dependent on the LF of the contaminant population. Therefore, even if some attempt is made to minimize the contamination (e.g., by introducing an LF prior into the photo-z calculation), the resulting LF is dependent on the LFs of all other galaxy populations at other redshifts.
\end{itemize}

Various procedures have been suggested for propagating the inherent uncertainties from photometric redshifts into the LF. A simple strategy, proposed by \citet{bolzonella}, is to carry out a Monte Carlo simulation, drawing redshifts for each galaxy at random from the probability distribution indicated by the photo-z algorithm. However, the fallacy in this approach is that the Monte Carlo redshifts are drawn from a probability distribution that is centered on the best-fit photo-z, and {\it not on the true redshift}. While this approach does indeed propagate the photo-z uncertainties into the final solution, it does not correct the systematic biases in the photo-z recovery; on the contrary, drawing galaxies from a photo-z probability distributions around photo-zs that are already affected by these uncertainties will convolve the solution even more, instead of deconvolving it, and may even exacerbate the systematic errors. A slightly different approach is taken by \citet{chen}, who incorporate the photo-z probability distribution into the likelihood function itself. However, inspection of their {\it ansatz} shows that the effect is, in fact, again a further convolution of the galaxy distribution, rather than a deconvolution, equivalent to the Monte-Carlo approach (it is equivalent to replacing each sample galaxy with a sum of ``fractional'' galaxies with a range of redshifts); it is thus a more elegant approach suffering from the same problem: the redshift probability distribution is evaluated around a redshift that is not the true redshift.

A mathematically much more
consistent method that addresses many of the criticisms above has been
suggested more recently by \citet{sheth}: The recovered photometric redshift is
treated as a quasi-observable distinct from the true redshift, and
reproducing its distribution is an additional requirement imposed on
the LF algorithm. This approach appears theoretically sound; however, it
stops short of replacing the photometric redshift with a different,
and easier-to-handle, observable, as we propose here. This raises two
practical problems for the application: 1) The algorithm is still based on the concept of a best-fit
photo-z. The procedure of first calculating a best-fit photo-z and
then feeding it as input into an LF algorithm always carries with it a
loss of information that is contained in the full photometric
vector. 2) The error function of the photometric redshift must be modeled well,
which is challenging, as it is a function of many parameters,
including the observed flux and true SED of a galaxy. The shape of the photometric redshift probability distribution function varies with the SED, redshift, and apparent magnitude of the galaxy in question (e.g., late-type galaxies are much more difficult to constrain than early-type galaxies), and may not even generally be assumed to be Gaussian either, but may in some cases exhibit a complex shape with secondary maxima.

An additional issue that has not been addressed by conventional LF
algorithms (but appears to be accounted for by \citet{sheth}) is that the {\it probability distribution} of photometric
redshift is itself dependent on implicit or explicit prior assumptions
about the LF of galaxies at any redshift. Photometric redshifts are
therefore dependent on prior assumptions about the redshift
distribution of galaxies (i.e., the LF of galaxies of all possible
spectral types throughout cosmic history). In photometric redshift
determinations, this dependence is often neglected, and redshifts are
calculated simply as best-fit estimates, regardless of the redshift or
template type that yields the fit. Note, however, that even this
approach to photo-z determination carries an implicit and unjustified prior, namely, that the prior probability distribution,
which depends on the cosmic volume and the LF, is flat with
redshift. A more promising approach is Bayesian redshift determination
\citep{benitez}, which applies priors (usually in the form of
luminosity function priors) to select the most probable photometric
redshift. However, while at least the shape of the bright end of the
LF is reasonably well known, priors about the faint end and the
high-redshift LF are highly uncertain. Furthermore, upon obtaining a
new solution for the LF at any redshift, the priors should be modified
to accommodate this new information, and the process iterated, so as to
avoid a contradiction between the LF result and the assumed
prior. If this is not done, a logical inconsistency between the prior
applied to the photo-z calculation and the resulting LF will generally result.

In summary, there are four points of criticism that may be raised against the methods employed to recover LFs from photometric samples in the past:
\begin{itemize}
\item Instead of a deconvolution, they have often carried out a further convolution of the data by applying an assumed error function to the observed data. A proper deconvolution requires that the error function be applied to a model, and the resulting distribution be compared to the observed data.
\item They have evaluated the predicted and observed galaxy
  distributions in a parameter space consisting of redshift and
  derived variables, which are not actual observables, but are
  affected by very complex error functions, which prohibit a direct
  deconvolution. In such cases, the comparison is more easily
  carried out in a parameter space consisting of direct observables,
  allowing us to convolve a predicted data set with the known error
  function instead.
\item They have used photometric redshifts and have consequently been forced to evaluate their error functions. Photometric redshifts have very complicated error functions that are usually neither Gaussian nor uniform, but instead feature secondary maxima and are dependent on magnitude and SED type. But the calculation of LFs does not require knowledge of the redshift of individual galaxies. We therefore suggest to eliminate the photometric redshift from the consideration and focus on direct observables, i.e., the observed fluxes, which have a much simpler error function.
\item They have often failed to take into account the dependence of the photometric redshift on prior assumptions about the LF at all redshifts. A proper determination of the LF must acknowledge this dependence and ideally provide a self-consistent solution that is not dependent on external priors, but constrained only by the data.
\end{itemize}

The problems described above arise from the attempt to use established
LF algorithms, which were originally developed for samples for which
the exact redshifts and, by extension, absolute magnitudes, are known,
in a context that they were not constructed for, namely,
photometric-only galaxy surveys. Our {\it Photometric Maximum
  Likelihood} method adopts the basic principle of maximum likelihood
algorithms: From a trial LF, the galaxy distribution in a certain
parameter space (e.g., absolute magnitude) is predicted, and the probability for drawing the observed sample from this distribution (i.e., the likelihood of the trial LF) is then calculated. The trial LF is then adjusted to optimize this likelihood. However, the {\it Photometric Maximum Likelihood} algorithm differs from established algorithms by comparing the observed and predicted galaxy distributions in a parameter space consisting only of direct observables, namely, the observed photometric fluxes in all available filter bands. We refer to this parameter space as {\it photometric space}. This addresses the problems described above.
\begin{itemize}
\item The PML is able to carry out a proper deconvolution of the observed photometric data: It applies an error function to the expectation values for the photometric fluxes in each filter band and compares the resulting distribution to the observed data.
\item By evaluating the galaxy distribution as a function of observed fluxes, instead of photometric redshifts, the PML avoids the unnecessary loss of information that is usually incurred when converting the full information of a multi-band photometric observation into a single photometric redshift. This also offers the decisive benefit that photometric errors in individual filter bands are much easier to model than those of photometric redshifts. 
\item The algorithm considers the contributions of all LFs at all
  redshifts to the distribution in photometric space simultaneously,
  and therefore finds a self-consistent solution for each LF that is
  (ideally) not dependent on implicit or explicit external luminosity priors.
\end{itemize}

Our algorithm, by design, naturally accounts for the dependence of the
LF on redshift and SED type. An important aspect is the fact that,
when evaluating the distribution of galaxies in photometric space, in
principle, any galaxy population at any redshift can contribute to the
galaxy counts at a given point in photometric space. It is not
possible to clearly separate samples by SED type and redshift. This
implies that the likelihood found by the PML algorithm is calculated
as a likelihood for all LFs of all galaxy populations throughout
cosmic history. This appears at first as a drawback, greatly enlarging
the number of free parameters involved in the problem. However, as has
been pointed out above, in the conventional approach, it is not
strictly possible to clearly separate a sample by redshift and SED
type either. Any association of an individual galaxy to a given
redshift and SED type is only a statistical one, and under proper
Bayesian consideration, the probability for any such association is
dependent on the prior assumptions about the LFs of all galaxy
populations that a galaxy could in principle have been drawn from. Whereas this
dependence usually goes unacknowledged, it is made explicit in the
PML, and the problem of introducing external priors is (in an ideal
case) eliminated by solving for all contributing LFs simultaneously.

\label{sec_principles}

\subsection{Derivation}

The derivation of a maximum likelihood algorithm starts with the definition of the likelihood function, i.e., the probability that the observed data of a galaxy survey would have been obtained from a given trial LF. The specific likelihood function that we use is inspired by the approach of \citet{marshall83}, which has also been used more recently by \citet{poli} and \citet{giallongo} for the determination of galaxy LFs, which uses a product of Poisson probabilities to derive the survey probability.

We imagine an $n$-dimensional flux space, the {\it photometric space}. In this space, each dimension corresponds to one filter band in which a flux is measured, and each coordinate to a measured flux in the corresponding filter. Each galaxy $i$ with measured photometry thus occupies a unique position $\vec{f_{i}}$ that represents the observed flux in all $n$ filters. We now imagine this space to be subdivided into small cells [$\vec{f}$;$\vec{df}$] that are chosen small enough that each cell contains either exactly 0 or 1 galaxies. We then define the likelihood function of the survey as the product of the Poisson probabilities that each cell that contains a galaxy would contain exactly one object, and that each of the remaining cells would contain none, under the prior of a chosen luminosity function. In other words,
\begin{equation}
\mathcal{L} = \prod_{i}^{N} \lambda(\vec{f_{i}}) \vec{df} e^{-\lambda(\vec{f_{i}})\vec{df}} \prod_{j} e^{-\lambda(\vec{f_{j}})\vec{df}} 
\label{eq_likelihood} 
\end{equation}

Here, $\lambda(\vec{f_{i}})\:\vec{df}$ is the expectation value of the
number of galaxies in cell [$\vec{f_i}$;$\vec{df}$] of photometric
space. The index $i$ runs over all observed galaxies, while the index
$j$ runs over all cells of flux space that are not occupied by a
galaxy. In contrast to \citet{marshall83}, \citet{poli}, and
\citet{giallongo}, we apply this likelihood evaluation to the
distribution of galaxies in photometric space, while the previous work
makes use of the availability of full or partial spectroscopic
coverage to evaluate the galaxy distribution as a function of
luminosity and redshift. This distinction is the defining feature of
the PML.

Recently, \citet{kelly} have suggested that the proper
statistical distribution to use is a binomial one, rather than the
Poisson distribution used in the {\it ansatz} of Marshall et
al. However, they acknowledge that, for surveys that sample only a
small fraction of the total sky, which is certainly the
case for MUSYC, Poisson distributions
are acceptable.

In our sample, we do not have the luxury of spectroscopic redshifts. It is therefore necessary to introduce a mapping between the luminosity function parameters and photometric space. We assume that the photometric data points are distributed according to a Gaussian distribution centered on the expectation value for a given absolute magnitude $M_0$, $SED$, and redshift $z_{0}$, and furthermore, that the photometric errors are independent in each filter.

 We do this by defining $\lambda$ in the following way:
\begin{eqnarray}
\lambda(\vec{f_{i}}) = \sum_{SED} \int dM_{0}\int dz_{0} \left(\frac{dV_{c}}{dz}\right) \: \frac{p_{\sigma}(\vec{f_{i}} \mid M_{0};SED;z_{0})}{\prod_n \Delta_n} \nonumber \\ 
\qquad \times \Phi(M_{0};SED;z_{0}\mid P) .
\label{eq_lambda}
\end{eqnarray}

This expression integrates over all possible galaxy populations --- at all redshifts, all absolute magnitudes, and all spectral types $SED$ --- and evaluates their contributions to the galaxy density in photometric space at the flux position $\vec{f_i}$. The term $dV_c/dz$ is the differential comoving volume, and $P$ are the assumed parameters of the LF whose likelihood we wish to test. The fraction of galaxies with the physical parameters $M_{0}$, $SED_{0}$, and $z_{0}$ that contribute to the galaxy density at a point $\vec{f_i}$ is determined via the probability term $p_{\sigma}$ --- the $\chi^{2}$ probability density resulting from a comparison between the fluxes predicted for these parameters, and the observed fluxes; this distribution is used here under the assumption that the errors in each filter band are uncorrelated with each other and that the photometric uncertainties in each band are Gaussian. This probability term is our assumed photometric error function. The denominator is the product of the observed flux uncertainties and serves to normalize the probability distribution. We use the following definition of the dimensionless probability density:
\begin{equation}
p_{\sigma}= \frac{1}{(2\pi)^{D/2}} e^{-\chi^{2}/2} .
\end{equation}
Here, $D$ is the number of degrees of freedom (in our case, the number of filters with valid photometry).

The $\chi^{2}$ quantifier is defined as
\begin{equation}
\chi^{2}=\sum_d^D \left( \frac{f^{i}_{d}-f^{pred}_{d}}{\Delta f^{i}_{d}} \right)^{2}.
\end{equation}

We use the {\it empirical} measurement uncertainties $\Delta f^{i}_{d}$ of galaxy $i$ to define the $\chi^{2}$ statistic. This is a compromise. A more consistent approach would be to select uncertainties appropriate for the {\it predicted} fluxes of the galaxy in each filter band. However, this would require an error model reproducing accurate uncertainties in every filter band as a function of flux. As the photometric uncertainties are largely dictated by conditions other than luminosity alone (such as the surface area of the galaxy), this would require additional assumptions that would weaken, rather than strengthen, our case. We believe that the substitution of the empirical uncertainties in this definition is a justifiable compromise.

A notable point about this approach is that the absolute magnitude of
an individual galaxy enters via the normalization of the SED in any
desired rest-frame filter band (which need not be identical to the
filters used in the survey). The $k$-correction is therefore integrated into the algorithm implicitly, and the absolute magnitude for a given galaxy at a given redshift and SED is constrained by all available filter bands, rather than by a single band and a corresponding $k$-correction; it is therefore usually quite precisely constrained. A caveat is that the accuracy of absolute magnitudes
depends on the accuracy of the templates; therefore, although the
algorithm is in principle capable of calculating LFs in any desired
rest-frame filter band, it is desirable that the luminosity in this
filter band be well-constrained by the available photometry (i.e., we
would distrust rest-frame LFs in filter bands at the extreme red or
blue end of the survey coverage).

As Eq. \ref{eq_lambda} shows, the PML accounts intrinsically for the dependence of the LF on redshift, spectral type, and, if desired, other parameters of the LF. It also automatically accounts for the effect of photometric uncertainties on the recovered LF. Most importantly, however, it combines the functionality of a photometric redshift algorithm with that of a LF algorithm via the probability term $p_{\sigma}(\vec{f_{i}} \mid M_{0};SED_{0};z_{0})$, which relates a given galaxy population to the galaxy distribution in photometric space. 

Equation \ref{eq_likelihood} can now be transformed by setting $S=-2\:ln\:\mathcal{L}$ (and writing $\lambda_{i}$ as shorthand for $\lambda(\vec{f_{i}})$):
\begin{equation}
S = - 2 \sum_{i}^{N} ln\lambda_i \vec{df} e^{-\lambda_i \vec{df}} -2 \sum_{j} ln\:e^{-\lambda_j \vec{df}}
\end{equation}

We split the argument of the first logarithm:

\begin{equation}
S = - 2 \sum_{i}^{N} ln\lambda_i \vec{df} -2 \sum_{i}^{N} ln\:e^{-\lambda_i \vec{df}} -2 \sum_{j} ln\:e^{-\lambda_j \vec{df}}
\end{equation}

\begin{equation}
S = - 2 \sum_{i}^{N} ln\lambda_i \vec{df} +2 \sum_{i}^{N} \lambda_i \vec{df} +2 \sum_{j} \lambda_j \vec{df}
\end{equation}

The second and third terms can be combined into an integral over all of observable flux space:
\begin{equation}
S = - 2 \sum_{i}^{N} ln\lambda_i \vec{df} +2 \int \lambda \vec{df}
\label{eq_l}
\end{equation}

Here, the first term represents the likelihood of drawing the observed
galaxies from the assumed input LFs, while the second term is simply
the total predicted number of galaxies within the flux limits of the
survey. The expression $\vec{df}$, the volume of one of our cells in
photometric space, can be omitted, as it will only add a constant
offset of $S$. Increasing the assumed space density of any galaxy
population (for example, by increasing the normalization of the LF)
will decrease the first term, reflecting the greater probability of
drawing the observed galaxies from such an input LF, but it will also
increase the second term, the total number of galaxies expected in the
survey. The luminosity function that balances these terms by minimizing the function $S$ is the maximum likelihood LF. 

Calculating $\lambda(\vec{f})$ in the second term requires an
assumption regarding the width of the distribution, i.e., the
photometric errors of a hypothetical galaxy at a point $\vec{f}$ in
photometric space, in order to correctly represent the contributions
of galaxy populations near the flux limit of the sample to the total
number of observed galaxies. It should be set to a value typical of
the survey under consideration; we have experimentally found that the
impact of this value on the recovered LFs is very minor. We set it to
$25\%$ of the flux, which is characteristic of
most objects near this flux limit. This value determines the fraction
of galaxies whose observed fluxes scatter across the survey limit, and
it will influence the solution only if the faint end of the LF is very
steep, i.e., if a much larger number of normally unobservable galaxies
scatter above the survey limits due to photometric errors than vice
versa. 

Despite our magnitude threshold of $R\leq25$, there are galaxies in
the catalogue that are consistent with zero flux in at least one
other filter band. As a result, the algorithm may associate them with
extremely faint populations that would normally not be expected to contribute
to galaxy counts at all, given our model SEDs and the nominal flux
error of 25\%. Inclusion of such objects may prevent a full
convergence of the LF solution, especially in parameters that are
constrained by few other objects. To prevent this, we only evaluate
contributions to $\lambda_i$ from populations that yield a non-zero contribution
to the total expectation value for the number of
galaxies in the survey. In other words, we ignore potential
contributions to the observed galaxy sample from populations that
should, assuming reasonable SEDs and photometric errors, be outside
the survey flux limits anyway. 

The fact that no precise redshift information is available does imply
that the PML shares a disadvantage with the traditional $V/V_{max}$
method: It requires certain assumptions about spatial homogeneity of the
LF, violation of which may bias the results. In a non-spectroscopic
sample, this is generally unavoidable, because photometric redshift
uncertainties wipe out detailed information about large scale
structure. To the extent that the photometric information permits it,
the PML may constrain the LF in several independent redshift bins and
thus reproduce gross features of the cosmic galaxy density fluctuation. 

\subsection{Parametrization of the LF}

A number of possible parameterizations have been proposed in the literature to represent the galaxy luminosity function. The classical representation is the Schechter function \citep{schechter76}, which describes the LF with a power-law faint end and an exponential cutoff at the bright end. It is characterized by three parameters: $M^{*}$, the characteristic bright-end magnitude, $\alpha$, the faint-end slope, and $\Phi^{*}$, a normalization parameter:

\begin{equation}
\Phi(M) = (0.4\:ln\:10)\:\Phi^{*}\:10^{0.4(M^{*}-M)(1+\alpha)}\:exp(-10^{0.4(M^{*}-M)})
\label{eq_schechter}
\end{equation}

As our method requires us to model the contributions from LFs of all galaxy populations at all redshifts to the observed galaxy distribution simultaneously, we follow \citet{giallongo} in using an additional parameter $\delta$ to model evolution in the parameter $M^{*}$:
\begin{eqnarray}
\Phi(M,z) = (0.4\:ln\:10)\:\Phi^{*}\:10^{0.4(M^{*}_{0}-\delta\:log_{10}[(1+z)/(1+z_0)] -M)(1+\alpha)} \nonumber \\
\qquad \times exp(-10^{0.4(M^{*}_{0}-\delta \:log_{10}[(1+z)/(1+z_0)]-M)}
\label{eq_evoschechter}
\end{eqnarray}

This equation is derived from Eq. \ref{eq_schechter} simply by
substituting $M^{*}$ at with a redshift-dependent $M^*(z)=M^*_0-\delta
\:log_{10}[(1+z)/(1+z_0)]$. Here, $M^{*}_0$ is the characteristic
bright-end magnitude evaluated at a fiducial redshift $z_0$. To
minimize covariances between $M^*_0$ and $\delta$, $z_0$ should be
chosen as the redshift at which $M^*$ is best-constrained by the
sample. In our case, we have experimentally found that, for both early
and late-type galaxies, the $M^*_0$-$\delta$ covariance is minimized
for $z_0\approx 0.95$.

For modeling the LF as a function of SED type and redshift, we assume
that each SED template type is described by a Schechter function with
a constant $\alpha$, $M^{*}$, and $\delta$. As photometric data cannot
resolve density fluctuation on arbitrarily small scales, we assume
that the LF for a given SED type is constant in each of ten redshift
bins that we divide the redshift range $0.02 \leq z < 6$
into. Potential density fluctuations on scales similar to or larger than these bins can
therefore by reproduced by the algorithm. The redshift bins, along
with the approximate limiting absolute magnitudes for early and late
type galaxies, are shown in Table \ref{tab_redshiftbins}. The
divisions between redshift bins are motivated by the desire to
apportion approximately equal numbers of observed galaxies to each
bin. Also, we have attempted to reflect the robustness of photometric
redshift constraints in the choice of binning. Although this model represents a compromise, the number of free parameters is still considerable.

Finally, to provide a visual verification of whether the Schechter function is indeed an adequate description of the overall shape of the LF, we provide a binned representation of the LF. Here the free parameters are simply the averaged values of the LF in a number of discrete bins: 

\begin{equation}
\Phi (M) = \sum_i \Phi_i\:H(M,i),
\label{eq_swlf}
\end{equation}

where $H(M,i)=1$ if $M_i\leq M < M_{i+1}$ and $H(M,i)=0$
otherwise. If evolution of the parameters $M^*$ and $\Phi^*$ is
suspected across the redshift range for which this representation is
used, the interval limits $M_i$ can of course be taken to be
relative to the evolving $M^*$ at a given redshift, and an additional
redshift-dependent normalization constant can be introduced to take
out any previously known redshift evolution of $\Phi^*$. This is the
approach that we will be following in \S \ref{sec_binnedlf}.

\begin{table}
\caption{Redshift bins}
  \begin{tabular}{@{}lccc}
  \hline
bin \# & z & M$_{lim}^{early}$ & M$_{lim}^{late}$ \\
\hline
1   & 0.02 -- 0.1 & -11.5 & -11.5 \\
2   & 0.1 -- 0.25 &  -14 & -14 \\
3  & 0.25 -- 0.4 &  -15.75 & -15.5 \\
4 & 0.4 -- 0.6 &  -17 & -16.5 \\
5  & 0.6 -- 0.8 & -18.5 & -17.75  \\
6   & 0.8 -- 1.2 &  -20 & -19 \\ 
7  & 1.2 -- 1.5 &  -21.5 & -20 \\
8 & 1.5 -- 2.5 &  -22.75 & -20.75 \\
9 & 2.5 -- 3.5 &  -23.75 & -21.75 \\
10 & 3.5 -- 6.0 &  -24.5 & -22.75 \\
\hline
\end{tabular} \\
Broad redshift bins within which we analyze the
  LF. M$_{lim}^{early}$ and M$_{lim}^{late}$ give approximate
  effective magnitude limits around which the completeness of our
  sample drops below $\sim 10\%$ for average early- and late-type SEDs.
\label{tab_redshiftbins}
\end{table}

\subsection{Error Calculation}
\label{sec_errors}
The PML returns a likelihood parameter $S = -2\:ln\:\mathcal{L}$ for
every set of trial LF parameters. Therefore, in principle, uncertainty
contours on any parameter can be determined with the method of \citet{avni}, i.e., the uncertainties are given by the set of parameter values of a constant likelihood
\begin{equation} 
\mathcal{L}=\mathcal{L}_{max} \times e^{-\Delta\chi^{2}/2} ,
\end{equation}
or, in terms of the likelihood parameter $S$ in Eq. , 
\begin{equation}
 S = S_{min} + \Delta\chi^{2} .
\label{eq_errors}
\end{equation}

Here, $\mathcal{L}_{max}$ is the maximum likelihood, and $\Delta\chi^{2}$ is the
$\chi^{2}$ value for the desired level of confidence and the
appropriate number of degrees of freedom. The latter is usually the
number of parameters whose uncertainties we are examining, i.e., two
if we are considering the uncertainties in the $M^{*}-\Phi^{*}$
plane. Following the example of Avni, we call these parameters the
``interesting parameters'', while referring to all other parameters
that the likelihood may or may not depend on, but whose values do not
interest us at the moment, as the ``uninteresting parameters''. 
Eq. \ref{eq_errors} requires that, for any choice of interesting
parameters whose likelihood we wish to probe in order to find their
uncertainty contours, the likelihood be maximized with respect to all
uninteresting parameters. 

There are some parameters in our model for the LF and its evolution
that we include neither in the best-fit search nor in the uncertainty
calculation; we refer to these as background parameters. In our case,
we treat the LF parameters of the very high and very low redshift LFs,
which are ill-constrained by our data and sensitive to systematic
errors, as such background parameters. We set these to
realistic values motivated by extrapolating the $\alpha$, $M^{*}$, and
$\Phi^{*}$ values from the constrained redshift bins in order to
represent the approximate contribution of the high- and low-redshift
universe to the observed galaxy counts. We do evaluate their impact on
our fits with a Monte Carlo approach by adding large Gaussian
perturbations to their values and repeating the best-fit search. This
approach is not meant to represent a realistic propagation of the
uncertainties of these high-redshift parameters, because neither are the
sizes of the propagated perturbations scientifically motivated, nor
are these parameters necessarily statistically independent. Rather,
this procedure demonstrates the covariances that exist between these
background parameters and the parameters constrained by our analysis,
and serves to illustrate whether a given recovered LF parameter
is robust against
such implicit or explicit assumptions about the LF in redshift regions
where it cannot be constrained directly.

\subsection{SED Fitting}
\label{sec_seds}
The Photometric Maximum Likelihood method for the calculation of the
LF, just like photometric redshift algorithms, requires a set of
spectral energy distribution (SED) templates for galaxies in order to
map the assumed LFs into a galaxy distribution in photometric space. The
templates we have chosen are based on the
empirical templates by \citet{cww}, extrapolated into the UV using
Bruzual-Charlot models \citep{bruzualcharlot}. These base templates correspond to Ellipticals
(E), Sbc galaxies, Scd galaxies, and very late-type irregulars
(Im). In addition, we use a starburst template (SB1) by
\citet{kinney}. We create linear combinations of
these templates and further apply a Calzetti dust extinction law
\citep{calzetti} with a range of extinctions to them, arriving at a
base template set comprising over one hundred SEDs. We then
optimize this template set with regard to its ability to reproduce
the redshifts of the subsample of galaxies with available
spectroscopic redshifts. The optimization is carried out by
determining for each possible template whether its inclusion in the
final template set would improve the quality of the best-fit
photometric redshift solution with regard to the $rms$ photometric
redshift residual and the fraction of highly discrepant
outliers. Templates are accordingly added to and removed from the set iteratively
until no further improvement is possible. Table
\ref{tab_templates} describes how the 12 final templates are generated from
the four CWW templates. One template, ``90\% Im + 10\% SB1 mod.'', has
been generated from a late-type one by strongly suppressing flux below
4000 \AA.

In the course of our analysis, we group
several of these templates into two broad classes, early and late
types. The possibility of a finer distinction exists, up to assigning
individual LFs to each SED. We have experimented with up to five
distinct SED classes, but have found the recovered LF shapes to fall
into two broad categories, one covering the two earliest-type
templates, and one covering all later-type templates. Therefore, we
will base our discussion of the LF results on just the two
aforementioned types. In the present analysis, each individual
template is weighted equally, i.e., every template belonging to the
same class (early or late) is assumed to be described by the same LF,
including the normalization. In subsequent work, we will explore the
option of weighting templates differently in order to reproduce a more
realistic representation of the distribution of the galaxy population in
photometric space.

Although we describe the 12 templates
using just two LFs, the large number of templates nonetheless has an
important effect: it broadens the footprint of a given LF in
photometric space, allowing it to account for galaxies that may be
incompatible with having been drawn from just any one individual
SED. Table \ref{tab_templates} shows how the templates are distributed into early and late types.

\begin{table}
\caption{Grouping of SED templates into SED classes}
  \begin{tabular}{@{}ll}
  \hline
 Template & Class \\
 \hline
 E                                  & early type \\
80\% E + 20\% Sbc                   & early type \\
Sbc                                 & early type \\
Scd                                 & late type \\
80\% Scd + 20\% Im                  & late type \\
80\% Scd + 20\% Im, E(B-V)=0.05 & late type \\
40\% Scd + 60\% Im                  & late type \\
30\% Scd + 70\% Im                  & late type \\
10\% Scd + 90\% Im                  & late type \\
Im                                  & late type \\
90\% Im + 10\% SB1                  & late type \\
90\% Im + 10\% SB1 mod.             & late type \\
\hline
\end{tabular} \\
The first column gives the linear combinations of the
five base templates E, Sbc, Scd, Im, and SB1, that we use in our
analysis. One template is slightly dust-reddened. The second column
shows how templates are assigned to the two SED classes, early and
late type.
\label{tab_templates}
\end{table}

An obvious caveat in this approach is that the empirical templates are
appropriate for bright low-redshift galaxies, but may not be representative
of the high-redshift universe, nor of very faint dwarf
galaxies. Furthermore, the empirical optimization of our
template set is naturally biased towards templates that represent the type
of galaxies in the spectroscopic subsample, which was not chosen
systematically. For these reasons, we take special care to verify the
robustness of our conclusions against assumptions about the
high-redshift LF (and thus, by extension, about the SEDs that we
associate these high-redshift LFs with).

In subsequent studies of the full MUSYC sample, we plan to address
these restrictions by allowing for a) evolution of the
template set with redshift and b) superpositions of templates to
overcome the discreteness inherent in the current template set. Both
approaches can be integrated naturally into the PML, as long as an
unambiguous assignment can be made between a certain SED template or
template combination and the LF that is supposed to describe this type
of galaxy.

\subsection{Verification of the PML with a Mock Catalog}

\subsubsection{Construction of the Mock Catalog}

We have carried out extensive tests of our algorithm to verify its ability to recover the LF and other physical parameters of the galaxies in our sample. Most of these tests are based on applying the method to a mock galaxy catalog. We use a mock catalogue extracted from a
$\Lambda$CDM numerical simulation
populated with GALFORM (version corresponding to Baugh et al., 2005)
semi-analytic galaxies.
The simulation consists of a total of $10^9$ dark-matter particles in a
cubical box of $1000$h$^{-1}$Mpc a side, followed from an initial
redshift $z=30$.  The background cosmology corresponds to a model with
matter density parameter $\Omega_m=0.25$,
vacuum density parameter $\Omega_{\Lambda}=0.75$,
a Hubble constant, $H=h 100$kms$^{-1}$Mpc$^{-1}$, with $h=0.7$,
and a primordial power spectrum slope, $n_s=0.97$.  The
present day amplitude of fluctuations in spheres of
$8$h$^{-1}$Mpc is set to $\sigma_8=0.8$.  This particular
cosmology is in line with recent cosmic microwave background
anisotropy and large scale structure
measurements (WMAP team, Spergel et al. 2007, S\'anchez et al., 2006).

The catalog provides, among other parameters, redshifts, $r$-band luminosities, and bulge luminosity fractions (i.e., the fraction of the total luminosity that is associated with a bulge component of the galaxies). The LF in the simulation output is independent of redshift. The simulation output is volume- and luminosity-limited and complete to $M_{r}=-16$ and z=3.

We have chosen a spectral energy distribution for each mock galaxy by
superimposing up to four different SED templates. The templates used
for this purpose are the four empirical templates by \citet{cww}, two
starburst templates by \citet{kinney} and the SDSS QSO template
\citep{sdssqso}. The templates are picked randomly, with a probability
distribution that varies depending on the bulge fraction of the input
galaxy. Therefore, the range of possible SEDs for the mock galaxies
far exceeds those represented by the 12 SED templates used in our
analysis, and the mock SED generally does not correspond exactly to
any one of the templates used for the LF recovery.  We then obtain
expectation values for the galaxy photometry in the $UBVRIzJK$-bands
by applying the transmission functions (filter curves, atmospheric
transmission, and quantum efficiency) of the MUSYC survey, while
simultaneously preserving the flux in the $r$-band specified in the
output of the cosmological simulation. Uncertainties are taken from
the MUSYC galaxy that best matches the expected fluxes under a
least-squares comparison.

\subsubsection{Photometric Redshift Recovery}

The PML, just like any photometric redshift algorithm, is based on the
assumption that the full range of physical characteristics of a galaxy
(redshift, spectral type, absolute magnitude) can be recovered from
the photometric properties of a galaxy. A simple way of demonstrating
that the data are of sufficient quality and that our code is capable
of recovering this information is to extract the information that
would normally be provided by a photometric redshift analysis, i.e.,
constraining $z$ and $SED$ through the photometric vector. This is
easily accomplished by finding, for each galaxy $i$, the model
parameters $z_0$, $SED_0$, and $M_0$ for which the likelihood term
$p(\vec{f_i}\mid z_0, SED_0, M_0)$ is maximized. This is equivalent to
the best-fit solution yielded by photo-z algorithms such as HYPERZ \citep{hyperz}.

\begin{figure}
\includegraphics[width=84mm]{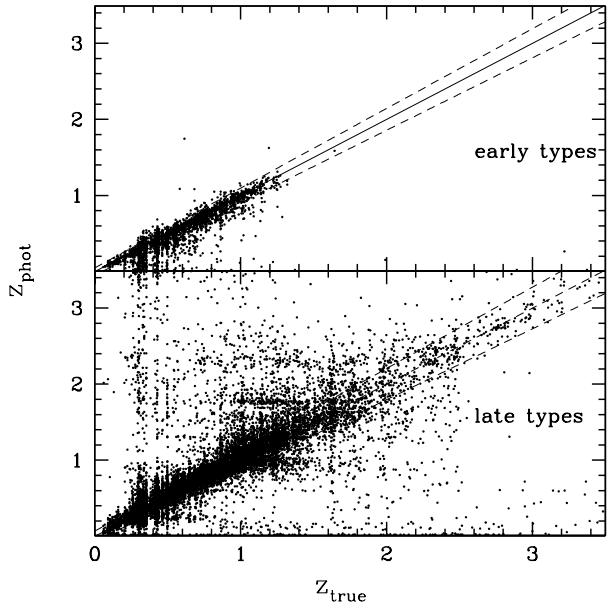}
\caption{Best-fit photometric redshifts recovered by our LF algorithm
from the mock galaxy catalogue.  Upper panel: Galaxies
best fitted with early-type templates. Bottom panel: Galaxies best
fitted with late-type templates.}
\label{fig_mockzz}
\end{figure}

Fig. \ref{fig_mockzz} shows the recovered photometric redshifts
versus the true redshifts in the input catalogue for 23256 mock
galaxies with $m_{R}<25$,
split by the recovered best-fit SED template type (early or late). The
figure shows a generally satisfying agreement between the input
redshift and the recovered photometric redshift, especially for
early-type objects. The photometric redshifts for late-type galaxies
are less-well constrained, presumably due to the lack of
characteristic structures such as the 4000 {\AA} Balmer break in
early-type galaxy spectra. There are preferred ranges for the
photometric redshift, clearly visible as horizontal bands in the
diagram, presumably indicative of redshifts at which characteristic breaks in the SED fall between neighbouring filters. It is worth pointing out that this is how purely Gaussian
errors in the photometry propagate, i.e., assuming a Gaussian
distribution for photometric redshift errors is, in general, a
fallacy.

The standard deviation of the redshift residual
$(z_{photo}-z_{true})/(1+z_{true})$ for all galaxies down
to $m=25$ is $\sigma=0.40$ (including
catastrophic outliers); when rejecting outliers of $3\sigma$ or more,
the residual standard deviation is $\sigma=0.059$ (with 13\% of the
objects having been rejected as outliers). The accuracy is slightly better for the early types ($\sigma=0.048$) than for the late types ($\sigma=0.061$). There is a slight systematic offset of $\Delta z/(1+z)=0.005$ (0.002 for early types and 0.005 for late types).

\subsubsection{Luminosity Function Recovery}

We will now demonstrate that our algorithm is able to recover the LF
from the mock galaxy catalog by carrying out three different analyses
of the mock catalogue. As the mock catalogue is populated using a
semianalytic galaxy formation model, and not a model Schechter
function, we need to recover parameters for the ``true'' LF from the
catalogue first. We do this by constraining mock galaxies to their
true redshift and the best-fit SED and absolute magnitude at that
redshift. In the second analysis, we will use the best-fit photometric
redshift,
SED, and absolute magnitude, equivalent to the conventional approach of substituting photometric for
spectroscopic redshifts. The third analysis
will consist of applying the full PML formalism to the mock
catalogue. 

The procedure for recovering the LF will be identical in all
cases. However, in order to constrain a galaxy to a given redshift,
SED type, and absolute magnitude, we modify Eq. \ref{eq_lambda} to
read
\begin{eqnarray}
\lambda_i = \sum_{SED} \int dM_{0}\int dz_{0}
\left(\frac{dV_{c}}{dz}\right) \: \delta(z_0-z_{true}) \nonumber\\
\times \delta(SED-SED_{true})\delta(M-M_{true}) \times \phi(M_{0};SED;z_{0}\mid P) ,
\label{eq_lambda_sty}
\end{eqnarray}
where $z_{true}$, $SED_{true}$, and $M_{true}$ are the values to which
a given galaxy is constrained; this approach is similar to the
original one by \citet{marshall83}. In the calculation of the second term
in Eq. \ref{eq_l}, we calculate $\lambda(\vec{f})$ as previously (an
integral over all of observable space), but apply a hard cutoff at the
survey magnitude limit without allowing for photometric
uncertainties. This procedure is then equivalent to running the
Marshall method on a parameter space comprising $z$, $M$, and $SED$,
rather than observed fluxes.

For each of the three analyses, we then follow the procedure that we
will subsequently apply to the real data: We model the LF as a
function of two SED types (early and late) in ten redshift bins. Over
the range $0.1\leq z < 1.2$ (redshift bins 2 to 6), we describe it in terms of
Eq. \ref{eq_evoschechter}, with a constant faint-end slope $\alpha$
and a two-parameter model for $M^{*}$ and its evolution with redshift
for each SED type. The normalization parameters $\Phi^{*}$ are
independent in all bins; $\Phi^*$ are included in the optimization in
redshift bins 7 and 8. All other LF parameters, specifically in the
lowest and highest redshift bins, are held constant during the
optimization, but periodically readjusted to match extrapolations from
the constrained bins.

An added complication is that the early-type LF in the input mock
catalogue is not a single Schechter function, but exhibits a possible
bimodality and a sharp cutoff at $M_r=-16$, neither of which is
represented by the LF models that we are fitting. However, this
complication will affect all three analysis methods.

\begin{figure}
\includegraphics[width=84mm]{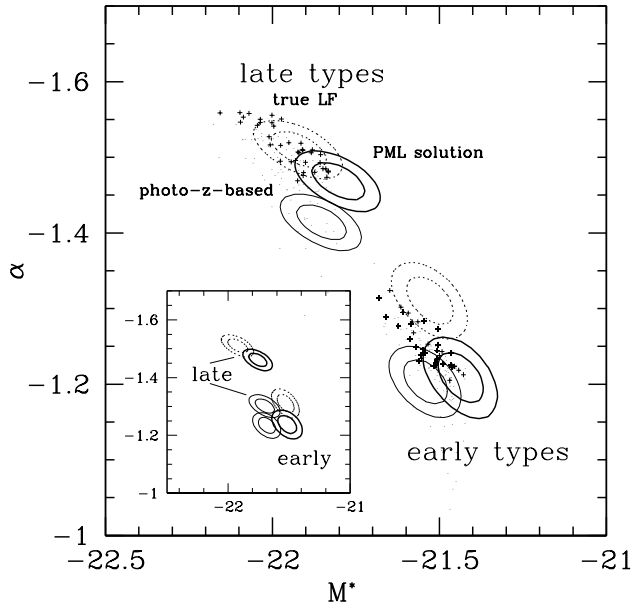}
\caption{Uncertainty contours at the 68\% and 95\% levels for the
LF parameters recovered from the mock catalogue using true
  redshifts (dotted contours), best-fit photo-zs (thin solid contours)
and the PML algorithm (thick solid contours). The main panel shows the
results from on $UBVRIzJK$ filter photometry, the inset only
$BVRIz$. Dots and crosses demonstrate how
  fluctuations in the high- and low-redshift LF parameters not
  included in the optimization may propagate into the recovered LF
  parameters, as described in the
  text (crosses symbolize a likelihood equal or better than the
  default scenario). 
 The photo-z-based LFs are inconsistent with the true LF, while the
PML solution overlaps with the true LF at at
least the $2\sigma$ level. The distribution of crosses shows that
different LF priors in the lowest- and highest-redshift bins might
bring the LF solutions in almost perfect agreement. The performance differences between
photo-z-based calculations and the PML are particularly striking for the smaller filter set.}
\label{fig_mock_calphams}
\end{figure}

Fig. \ref{fig_mock_calphams} shows the 68\% and 95\% error contours in the
$\alpha$-$M^*_0$ plane for all three analyses: the calculation based
on the true redshifts in dotted lines, the calculation using best-fit
photo-zs in solid thin lines, and the PML solution in thick
lines. Both for early and late type populations, the conventional
approach, using best-fit photo-zs, fails to recover the correct LF
parameters in the $M^{*}-\alpha$ plane. Although the fits would appear
consistent if projected only onto the $M^{*}$ axis, a comparison
involving only $M^{*}$ must be carried out at the same
$\alpha$, in which case the discrepancy between the photo-z-based
solution and the true LF is even more apparent.

The late-type LF parameters recovered by the PML, on the other hand,
are consistent with the solution obtained from the true redshifts. For the early-type population, there is a discrepancy in $\alpha$, so that the 2$\sigma$ contours barely
overlap; however, for an $\alpha\approx -1.25$, which is consistent
with both the PML analysis and the true redshift calculation, the
recovered $M^{*}$ are also perfectly consistent, unlike the conventional,
photo-z-based solution, which is significantly brighter. 

The PML solution can be brought into even better agreement for
different choices of values for the constant background parameters at
very low and high redshift. We have carried out the procedure
described in \S \ref{sec_errors} to determine how uncertainties in the
LF priors in the very high- and low-redshift bins propagate into the
LF fit. For each trial calculation, we apply Gaussian
perturbations of order 1 mag to the M$^{*}$ background parameters
(i.e., those not included in the optimization), and of order unity to the quantity $log_{10} \Phi^*$ (i.e., $M^{*}$ typically scatter randomly within 1 mag, and $\Phi^*$ within a factor
of 10 from the default values, which are very generous assumptions). We then calculate the best-fit
solution for the constrained parameters.

The best-fit results of these Monte Carlo trials are shown in
Fig. \ref{fig_mock_calphams} with dots (for all fits) and crosses (for
those fits that yield an equivalent or better likelihood after
optimization than our default scenario). Each symbol represents one
Monte Carlo trial with a different LF prior in the low- and
high-redshift bins, after the maximum likelihood optimization of the
free parameters has been carried out.

A large fraction of the Monte Carlo trials yield a better total
likelihood than our default fit. Since we do not apply any prior to judge which of the various
 background LF parameters that are probed by the Monte Carlo algorithm
 are realistic, and because, for reasons discussed in greater detail
 in \S \ref{sec_cosmiclf}, we do not consider our data set to be ideally suited for constraining the
high-redshift LF, we do not actually prefer any of the Monte Carlo
trials with better likelihood to our default fit; however, regardless
of whether the background LFs assumed for the Monte Carlo trial is
realistic or not, runs that yield a best-fit likelihood that is
significantly worse than our default scenario can be ruled out;
therefore, only the crosses in Fig. \ref{fig_mock_calphams} should be considered.

We find that, of the Monte Carlo trials that result in an equivalent or
better likelihood than the default fit, almost all yield LF parameters
that are in even better agreement with the true LF; in fact, for some
background priors, we recover LF parameters that are in almost perfect
agreement with the input LF (we have not introduced criteria, however,
to decide whether the background LF priors corresponding to these
trials are physically plausible or not, which is why we do not elect any of these
as the real best fit). Conversely, of those LF parameters that
are in worse agreement with the true LF than the default fit, almost
all are marked by having a substantially worse likelihood and are thus
ruled out. This means that the PML still has some constraining power in
redshift regions that we have chosen not to include in our
optimization in our conservative approach. 

However, the Monte Carlo approach does not open up additional degrees
of freedom to the photo-z-based algorithm, because (at least if the
photo-zs are derived from frequentist best-fit estimates), each galaxy
is uniquely associated with a given redshift bin, independently of the
LF assumed for it. Therefore, the photo-z-based LF remains clearly
inconsistent with the true LF.

The advantage of the PML method over the use of photometric redshifts
is even more obvious if we reduce the number of filter bands. Using
only the $BVRIz$ filters, we recover the results shown in the inset in
Fig. \ref{fig_mock_calphams}: Here, the photo-z-based results are
drastically different from the ``true'' LFs; the late-type LF is found
to be $\alpha\approx -1.3$, compared to the actual $\alpha=-1.52$, and
the recovered early-type LF is also significantly brighter and
shallower than the input LF. In contrast, the uncertainty contours of
the PML solution again overlap with those of the ``true'' LF.

\section{Results and Discussion}
\label{sec_results}
\label{sec_discussion}
\subsection{Magnitude Limits}

\begin{figure}
\includegraphics[width=84mm]{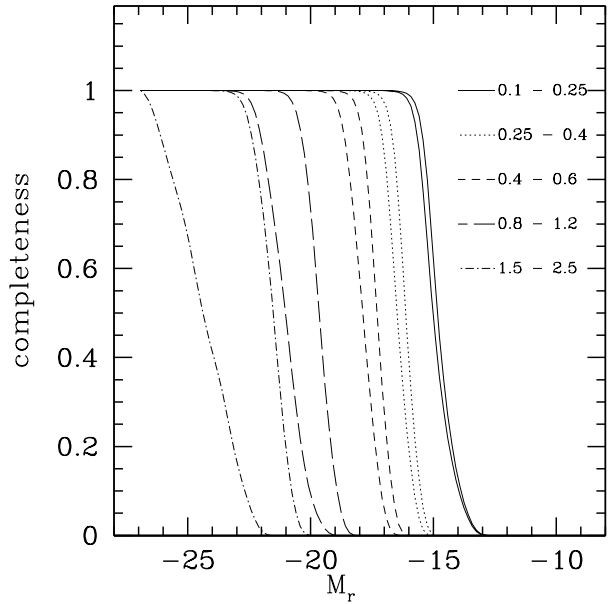}
\caption{Effective $r$-band magnitude limits for five of our ten redshift bins. Different line styles indicate different
  redshift bins. For each redshift bin, we draw two completeness
  curves, corresponding to very late and very early types. Assignment
  of line styles to redshift ranges is given in the plot.}
\label{fig_maglimits}
\end{figure}

How deep does MUSYC probe the galaxy LF in our individual redshift
bins? To answer this question, we plot in Fig. \ref{fig_maglimits} the derivative of the second term of the
likelihood function in Eq. \ref{eq_l} with respect to the value $\Phi(M)$ of the
LF at a given absolute magnitude; this is, in essence, the volume over
which a galaxy with a given SED and absolute magnitude in a given
redshift bin would be visible. It has here been normalized to unity
for the brightest galaxies. The figure shows the completeness for
the earliest- and the latest-type SED in redshift bins 2 ($0.1 \leq z
< 0.25$), 3 ($0.25 \leq z < 0.4$), 4 ($0.4 \leq z < 0.6$), 6 ($0.8 
\leq z < 1.2$), and 8 ($1.5 \leq z < 2.5$).

 For higher redshifts, the early- and late-type completeness curves
 differ significantly, indicating that late-type galaxies are more
 easily detectable than early-type ones. The curves do not cut off
 abruptly at a given absolute magnitude due to the fact that each
 redshift bin has a finite extent in redshift; galaxies near the
 low-redshift boundary of a bin are visible to fainter absolute
 magnitudes than those at the high-redshift boundary. In the
 lowest redshift bin under analysis here, our sensitivity extends to $M_{R}\approx-14$.

\subsection{Photometric Redshift and Luminosity Recovery}

We have already demonstrated that the PML is capable of recovering the
redshift of galaxies in the mock catalog. In the application to real
data, additional systematic errors that are not modeled by the mock
catalog may come into play, such as systematic biases in the
photometry that our zero-point corrections may not have
sufficiently compensated for. The best-fit photometric redshifts recovered by
our algorithm are a convenient way of testing for the presence of such
serious systematic errors. For this comparison, we have drawn
spectroscopic redshifts in the E-CDFS field from various public and
proprietary sources, including the NASA Extragalactic Database (NED),
publications by \citet{ravikumar} and \citet{vanzella}, as well as
redshifts determined within the MUSYC project, yielding a total of
2602 redshifts.

\begin{figure}
\includegraphics[width=84mm]{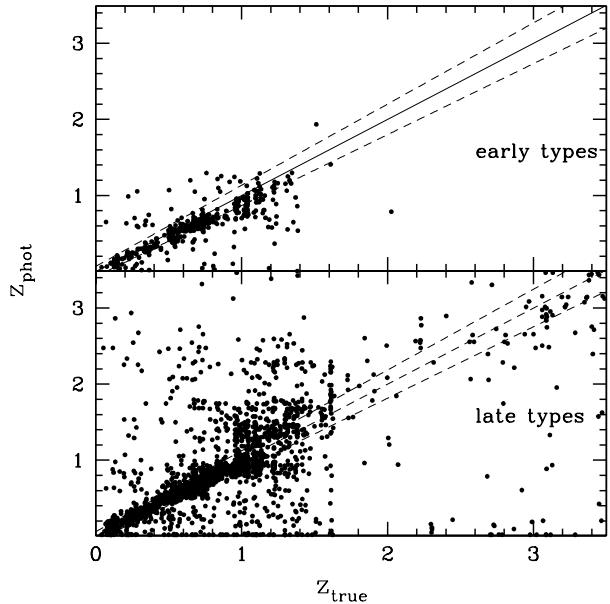}
\caption{Best-fit photometric redshifts compared to known
spectroscopic redshifts in the MUSYC catalog. Upper panel: Galaxies
best fitted with early-type templates. Bottom panel: Galaxies best
fitted with late-type templates.}
\label{fig_zz}
\end{figure}

Fig. \ref{fig_zz} shows the performance of the recovered best-fit
photometric redshifts versus spectroscopic redshifts. The plot
resembles our previous comparison using the mock catalogues in
Fig. \ref{fig_mockzz}: The performance is generally good; late-type
objects suffer from more catastrophic failures. The standard deviation of the redshift residual
$(z_{photo}-z_{true})/(1+z_{true})$ is $\sigma=0.27$ (including
catastrophic outliers); when rejecting outliers of $3\sigma$ or more,
the residual standard deviation is $\sigma=0.062$ (with 17\% of the
objects having been rejected as outliers). The accuracy is similar for early types ($\sigma=0.061$) and for late types ($\sigma=0.063$).
For comparison, prior to applying our photometric corrections to the zero-points and
uncertainties, these numbers are $\sigma=0.32$ and $\sigma=0.082$,
respectively ($\sigma=0.077$ for early types and $\sigma=0.083$ for late types). 

 Comparison between Fig. \ref{fig_mockzz} and
Fig. \ref{fig_zz} shows that many of the structures (particularly
the horizontal patterns showing preferred values for photometric
redshifts) found in the real mock data are visible in the observed
data as well. This provides important verification that the pattern of
catastrophic failures in the photometric redshift recovery can be
reproduced on the basis of just the Gaussian errors in the
input photometry and that, by extension, such color degeneracies that
normally present a grave problem for photo-z based LF determines can
be accounted for naturally by the PML. 

 However, there is an indication of photo-zs of early type galaxies being systematically
underestimated at $z>1$; this feature is only partially reproduced by the
mock data, and thus not fully corrected for by the
PML either. Therefore, we restrict our analysis to the redshift range $0.1
\leq z < 1.2$, the last redshift bin covering the fairly wide range
$0.8 \leq z < 1.2$.

The fact that our photometric data and SED template set allow us recover photometric redshifts with acceptable accuracy is encouraging, even though our algorithm does not use individual photometric redshifts per se. For the purpose of calculating luminosity functions, the ability to recover the absolute magnitude of a galaxy and account for $k$-corrections and other cosmological effects is at least as important. Although, unlike the redshift, the absolute magnitude cannot be verified by alternate observations, the COMBO-17 survey \citep{wolf2004} offers a high-quality comparison catalogue that covers the same field and contains estimates of absolute magnitudes in the
$r$-band. Crossmatching the COMBO-17 catalogue and the MUSYC catalogue yields 26234 galaxies. After adding an offset of $0.055$ mag to the COMBO-17 $R$-band magnitudes to convert from the Vega to the AB system \citep{freigunn94}, the apparent magnitudes in both catalogues agree well apart from a systematic
offset of $m_{MUSYC}-m_{COMBO-17}\approx +0.045$ (in the $R$-band; total
magnitudes for COMBO-17, and either corrected aperture or AUTO
magnitudes for MUSYC, as described in \S \ref{sec_data}). 

To evaluate our ability to recover absolute magnitudes from the MUSYC photometric data, we now constrain the MUSYC galaxies to be at the photometric redshift listed in the COMBO-17 catalogue, and carry out an SED fit at that redshift to calculate the $r$-band rest frame absolute magnitude. For a given redshift and SED type, the PML calculates the rest-frame luminosity of a galaxy directly from
the normalization factor that yields the best fit of the SED to the
observed fluxes. We compare these absolute magnitudes to the $r$-band
estimates provided by \citet{wolf2008} by calculating the quantity
$(M_{MUSYC}-M_{COMBO-17})-(m_{MUSYC}-m_{COMBO-17})$, which accounts
for the measured differences in observer-frame $R$-band apparent
magnitudes to isolate only the differences in the recovered absolute
magnitude. Offsets between the AB and Vega systems are negligible in
the rest-frame $r$-band \citep{freigunn94}. We restrict this
comparison to galaxies with $R<22$, because fainter galaxies ($R>22.5$) in the
COMBO-17 catalogue exhibit a strong quantization in the best-fit photometric
redshifts that casts doubts on their reliability. We then find a mean
of $\sim -0.06$ with a dispersion of $0.22$ over the redshift range
$0.1\leq z < 1.0$, i.e., MUSYC galaxies are inferred to be slightly more
luminous at the same redshift and for the same rest-frame $R$-band
magnitude. This value is only very weakly dependent on redshift (with a
slope of $-0.05$ and a zero point of $-0.04$ at z=0). A small
contribution to this discrepancy may come from differences in the
cosmological model in the calculation of absolute magnitudes
($\Omega_{\lambda}=0.73$ and $\Omega_m=0.27$ in our study,
$\Omega_{\lambda}=0.7$ and $\Omega_m=0.3$ for COMBO-17); this
difference is $\sim 0.04$ mag at $z=1$ in the sense of MUSYC galaxies
being inferred to be brighter. Therefore, if residual systematic
discrepancies exist in our calculation of absolute rest-frame
magnitudes in comparison to COMBO-17, they are at the level of $<0.1$ mag for the redshift range that we are analyzing. Possible
reasons for this residual discrepancy include the fact that our
rest-frame luminosities are constrained by all available filter bands,
so that the subtraction of the $(m_{MUSYC}-m_{COMBO-17})$ term in the
$R$-band will not account for all discrepancies in the input
photometry, as well as the different template set.

\subsection{The Cosmic LF}

\label{sec_cosmiclf}

We now apply the PML to the MUSYC data to calculate the field galaxy
 LF. For this purpose, we use the parametrization given by
 Eq. \ref{eq_evoschechter}, i.e., for each of the two SED classes, we
 assume a characteristic bright-end magnitude $M^{*}_{0}$ at a
 fiducial redshift $z=0.95$ (experimentally found to be the redshift
 for which $M^*$ is constrained best in this sample), a faint-end
 slope $\alpha$, and an evolutionary parameter $\delta$ to describe
 the variation of $M^{*}$ with redshift according to
 $M^{*}=M^{*}_0-\delta\times lg[(1+z)/(1+z_0)]$. The normalization parameters
 $\Phi^{*}$ are independent for each redshift bin and SED class. For the present work, we have chosen to calculate the LFs in the rest-frame $r$-band. This band is less sensitive to effects of current star formation than, for example, the $B$-band, but at the same time is well-constrained by the available photometry over the redshift range of particular interest to us.

The purpose of this calculation is to recover an approximate
representation of the LF, albeit with a low number of free parameters,
which will serve as the starting point for a subsequent refinement
with a more detailed {\it ansatz}. Furthermore, we will use this
calculation to constrain the faint end slopes $\alpha$ for the early and late SED types from
this calculation and adopt them for the rest of the analysis. While it would be possible to optimize with respect to the faint end slopes $\alpha$ in individual redshift bins, the flux limit prevents us from obtaining useful constraints on $\alpha$ in any but the lowest-redshift bins, up to $z\approx 0.4$ --- $0.5$. A follow-up study of a larger sample will explore this issue.

Parameters that we include in the optimization are M$^{*}_0$, $\alpha$,
and $\delta$ for both SED types, describing the shape of the LF over
the range $0.1 \leq z < 1.2$. Furthermore, we constrain $\Phi^{*}$
over the range $0.1 \leq z < 2.5$. All other LF parameters (i.e.,
$\alpha$, $M^*$ in redshift bins 1 and 7 to 10, and $\Phi^{*}$ in bins
1, 9, and 10) are held constant. The decision to treat the
high-redshift LF parameters as constant background parameters is based
on a number of criteria:
\begin{itemize}
\item The rest-frame $r$-band is not constrained directly by the available
  filter complement at high redshift.
\item Only the tip of the LF is sampled, which may introduce strong
  systematic errors if the most luminous galaxies do not obey the same
  Schechter function implied for the bulk of the galaxy population.
\item The applicability of z=0 empirical templates to high-redshift
  galaxy populations is doubtful.
\item Photometric redshifts, using this catalogue and our template
  set, have proven much less reliable at $z>1.0$, as is also shown by Fig. \ref{fig_mockzz} and Fig. \ref{fig_zz}.
\end{itemize}

  At low redshift, on the other hand, constraints on the LF are weak
  due to the small number of galaxies and the large uncertainties
  affecting the
  absolute magnitudes of individual galaxies.

The initial values that we choose for these background parameters are motivated by
the results of \citet{giallongo}. Their work uses two different
criteria to separate early and late type galaxies --- colors and
specific star formation rate ---, and examines two ways of defining
the threshold values --- empirically, from the observed bimodality, or
from a theoretical evolutionary track. This yields four different
estimates for the LF and its evolution for a given galaxy type. To
obtain an estimate in a given redshift bin, we evaluate their fit at
the central redshift of the bin, and average the four extrapolated
values for $M^{*}$ and $\Phi^{*}$. To convert their $B$-band LFs into
our $r$-band rest-frame absolute magnitudes, we adjust $M^{*}$
by $B-r=1.35$ for early types and $B-r=0.9$ for late types. These
values correspond to typical $B-r$ colors for early and late
populations from \citet{wolf} at low redshift. We caution that the
distinction made by Giallongo et al. between early and late types is different from ours; in
particular, while we use redshift-zero SED templates to
fit early and late type LFs, their threshold values for parameters
such as color and specific star formation rate vary with redshift. We therefore expect that ``early'' types would be more
prevalent at high redshift according to their definition than to
ours. 

We repeat the best-fit search several times; after each iteration, we
adjust the background parameters by averaging their previous values
with linear extrapolations of the best-fit $M^{*}$ and $\Phi^{*}$
values over $lg(1+z)$. Above the last redshift bin directly
constrained by our fit (above z=2.5), we cap the extrapolated values
for these parameters at those found in the closest redshift bin where
they are constrained directly, so as to avoid
implying unrealistically bright or abundant galaxy populations in the
high-redshift universe.

We solve for the maximum-likelihood solution in the remaining
16-parameter space (2 SED types $\times$ (3 shape parameters + 7
normalization parameters)) using an iterative algorithm that, starting
from the previous solution, first searches for the maximum likelihood point along each parameter
axis and thus deriving a shift $\Delta p_i$ in each parameter
$p_i$. Subsequently, it applies a shift $x \vec{\Delta p}$ to
all parameters simultaneously and calculates the likelihood as a
function of the scaling factor $x$ in the same manner. If the resulting adjustment in
any given parameter is much smaller than the spacing between the 
trial points that are initially evaluated along the parameter axis,
this spacing is slightly reduced. The algorithm stops when the trial
point spacings along all parameter axes are much smaller than the typical statistical uncertainties.

\begin{figure}
\includegraphics[width=84mm]{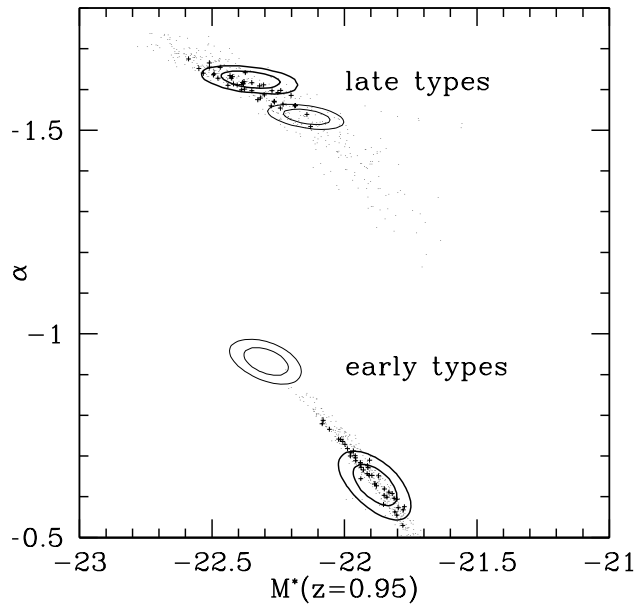}
\caption{Uncertainty contours for the best-fit LF parameters at 68\% and 95\% confidence
  levels in $\alpha$-$M^{*}$ parameter space. Bold contour lines show
  the PML solution, while thin contour lines indicate a solution based
  on a conventional algorithm using best-fit photo-zs. The results
  reflect a dichotomy between the early-type LF with a shallow
  faint-end slope and the late-type LF with an extremely steep slope
  of $\alpha\approx-1.6$. Dots and crosses again demonstrate how
  fluctuations in the high- and low-redshift LF parameters propagate into the recovered LF
  parameters. Only trial runs symbolized with crosses yield an acceptable likelihood.}
\label{fig_calphams}
\end{figure}

The best-fit values for the 16 parameters included in the optimization
are given in Table \ref{tab_cosmiclf}, along with the values adopted for the constant ``background'' parameters. The projection of the $\alpha$-$M^{*}_0$ plane is shown in
Fig. \ref{fig_calphams}, along with the 68\% and 95\% error
contours. Here, bold error contours indicate the results of the PML
analysis, while thin lines represent the classical calculation using
fixed photo-zs; the uncertainty regions are clearly disjunct.

The qualitative behaviour of the fits is consistent with
many previous literature results (i.e., \citet{bromley,wolf}) regarding the late- and early-type LFs, whose most
important characteristics are a rather shallow slope for early-type
galaxies and a significantly steeper slope for late-type
galaxies. However, the steepness of the slope of our late-type LF is
remarkable indeed; we find $  -1.62^{+   0.02}_{-   0.02}$,
compared to $  -0.62^{+   0.04}_{-   0.06}$ for the early types. Literature values for the
total LF tend to prefer values of $\alpha=-1.2$ to $-1.3$. However, it is
also known that the late-type LF is steeper than this average, and
\citet{blanton2005} find $\alpha=-1.3$ in the SDSS, but speculate
(based on the fact that the SDSS is likely to miss many low-surface
brightness objects) that it may be as steep as $-1.5$ or even
steeper. Such steep slopes at the faint end can be easily reconciled with older values of
$\alpha\approx-1.2$ by noting that the bright end of the LF is
dominated by early-type galaxies, and that the late-type LF with its
steep slope will begin to dominate the total LF at several magnitudes
fainter than $M^{*}$, i.e., only surveys combining very deep samples
with sufficient statistics to constrain the extreme faint end will be
able to identify this as an upturn. It is conceivable that this
dichotomy may also be identified as a ``dip'' in the overall LF,
especially in denser environments, where the steep late-type LF is
likely to have been depleted by environmental processes and its
normalization $\Phi^{*}$ lowered. Indeed, claims of such a ``dip'' feature in
the overall LF have been raised in the past based on surveys of galaxy groups
\citep{flint,miles2006,mendesdeoliveira06} and the Coma cluster \citep{biviano96}.

The early and late type populations show a very different evolutionary
behaviour with redshift. For the early-type sample, we find
$\delta\approx 0$, i.e., no evidence for luminosity evolution up to
$z=1.2$. In contrast, the late-type sample is fitted by an evolving LF
with $\delta\approx 2.5$, i.e., $M^{*}$ brightens by about 0.75 mag
from $z=0$ to $z=1$. However, the uncertainties on the $\delta$
parameters are very large, and even the late-type sample is formally
consistent with very weak luminosity evolution.

Our two-parameter fit to $M^*(z)$ can be extrapolated to $z=0$ to
recover the local LF parameters; however, as $M^*$ is primarily
constrained by galaxies at $z\approx 0.95$, uncertainties in $\delta$
propagate strongly into the extrapolated value. Formally, we find
$M^*(z=0)=-21.854^{+  0.200}_{-  0.200}$ for the early type and
  $M^*(z=0)=-21.632^{+  0.340}_{-  0.400}$ for the late type
    population. However, better constraints on the LF at z=0 can
    be obtained by fitting the LF parameters directly in individual
    low-redshift bins, and we refer the reader to the next section, \S
\ref{sec_individualzbins}, where we will discuss this question in
greater detail.

As we have discussed in \S \ref{sec_errors}, we presume LF parameters at very low
and very high redshift to be constant instead of constraining them directly from the data, as they are very susceptible to
systematic errors. However, as LF parameters in different redshift
bins are usually covariate, the exact LF solution recovered here
depends on our assumptions for the values of these background
parameters. We have applied the Monte Carlo approach described in \S
\ref{sec_errors} to gauge the sensitivity of the best-fit parameters
to these background values. For each trial calculation, we apply Gaussian
perturbations of order 1 mag to the M$^{*}$ background parameters, and
of order unity to the quantity $log_{10} \Phi^*$ in these background
bins (i.e., $M^{*}$ typically scatter randomly within 1 mag, and $\Phi^*$ within a factor
of 10 from the default values). We then calculate the best-fit
solution for the constrained parameters. The results are shown in
Fig. \ref{fig_calphams} in the M$^{*}_0$-$\alpha$ plane. Dots
represent Monte Carlo runs for which the final, best-fit likelihood
is substantially worse than for our default values; this means that
the corresponding configuration of background parameters is ruled out
by the data, and the run can be discarded. Crosses represent those runs for which we
achieve a similar or better maximum likelihood than for our default
background choices; these represent background configurations that
cannot be ruled out by the data, although we do not judge whether the
configuration of background LFs assumed in each Monte Carlo trial is
realistic, and therefore do not prefer any of these solutions to our
default scenario. According to Fig. \ref{fig_calphams}, variations in the background
parameters may then introduce a scatter that exceeds the 2$\sigma$
uncertainty contours. However, the qualitative picture remains
unchanged; in particular, none of the acceptable Monte Carlo trials
yield $\alpha>-1.5$ for the late-type population. Furthermore, only
very few outliers in the Monte Carlo analysis yield LF parameters
consistent with the photo-z-based calculation for the late-type
population, and none of them are consistent with the photo-z-based
early-type LF. We furthermore point out that the amount by which the
background parameters are allowed to scatter from the default
assumption is very generous. We interpret this as indicating that, to
change the faint-end-slope $\alpha$ of the late-type population to
values shallower than $-1.5$, $\Phi^{*}$ at high redshift would have to be permitted to
deviate by more than a factor of 10 upward from the values found at
lower redshift, or $M^{*}$ by more than a magnitude towards the bright
end, i.e., the high-redshift galaxy population would have to be much
brighter and more abundant than at $z<1.2$, which is a scenario that
we consider unrealistic.

In order to test the effect of excluding detections with questionable
star/galaxy classifications, we have
carried out an alternate fit including only objects with
$S/G\leq0.8$ (as opposed to our default of 0.97). The resulting best-fit values
differ slightly from those obtained above; most notably, $M^{*}_0$ for
the early types is fainter at $M^{*}_0 = -21.62$, and the
faint-end slope is slightly shallower at $\alpha=-0.43$. However,
there is virtually no change in the late-type LF, where the best-fit
values are $M^{*}_0=-21.66$ and $\alpha=-1.61$. As $\alpha$ and $M^{*}_0$
are correlated, it is not immediately clear whether the shift in the
early type best-fit parameters is due to a change at the bright or the
faint end; however, as there is no obvious tendency for objects with
$S/G>0.8$ to be associated with unusually high luminosities, we
suspect that the change is due to the removal of compact sources with
photometric properties characteristic of low-redshift early-type
objects. We return to this question once more in \S \ref{sec_binnedlf}, where we analyse the shape of the LF in a binned representation.

Just as doubts regarding the appropriateness of our empirical SED
templates prevent us from constraining the LF in the high-redshift
universe, we may also wonder whether a mismatch between our late-type
templates and the photometric properties of dwarf galaxies may have
steepened the observed faint end slope artificially. Dwarf galaxies,
typically being of low metallicity, may have significantly different
spectral properties than the brighter objects that the CWW templates
were originally drawn from. However, we would expect template
insufficiency to {\it lower} the recovered LF value associated with
the galaxy populations that are not represented well by the template,
while the LF of other populations with similar photometric properties
would be boosted instead. Therefore, a greater concern regarding the
steep faint end lies with the question whether the high-redshift LF is
represented well by our templates. The Monte Carlo procedure described
above, however, demonstrates that the steep faint end slope is not
very sensitive to the LF values assumed for the high-redshift population and cannot be pushed to values shallower than
$\alpha\approx-1.5$ even for the rather large background fluctuations
assumed there.

\begin{table*}
\begin{minipage}{110mm}
\caption{LF Results at $0.1 \leq z < 1.2$}
  \begin{tabular}{@{}llrrrr}
  \hline
bin & z & $\alpha$ & M$^{*}$ & $\delta$ & lg $\Phi^{*}$ \\
\hline
&early types \\
... & ...  & $  -0.62^{+   0.04}_{-   0.06}$ & $-21.885^{+  0.075}_{-  0.075}$ & $  0.107^{+  0.780}_{-  0.720}$ & ... \\
 1 & 0.02 -- 0.10 &  -0.639 & -21.918 & ... &  -2.502 \\
 2 & 0.10 -- 0.25 & ... & ... & ... &  $   -2.321^{+   0.043}_{-   0.037}$\\
 3 & 0.25 -- 0.40 & ... & ... & ... &  $   -2.726^{+   0.044}_{-   0.039}$\\
 4 & 0.40 -- 0.60 & ... & ... & ... &  $   -2.632^{+   0.034}_{-   0.026}$\\
 5 & 0.60 -- 0.80 & ... & ... & ... &  $   -2.482^{+   0.025}_{-   0.020}$\\
 6 & 0.80 -- 1.20 & ... & ... & ... &  $   -2.756^{+   0.028}_{-   0.027}$\\
 7 & 1.20 -- 1.50 &  -0.639 & -21.867 & ... &  $   -2.709^{+   0.053}_{-   0.056}$\\
 8 & 1.50 -- 2.50 &  -0.639 & -21.849 & ... &  $   -2.685^{+   0.077}_{-   0.079}$\\
 9 & 2.50 -- 3.50 &  -0.639 & -21.828 & ... &  -2.850 \\
10 & 3.50 -- 6.00 &  -0.639 & -21.802 & ... &  -2.953 \\
\hline
&late types\\
... & ...  & $  -1.62^{+   0.02}_{-   0.02}$ & $-22.345^{+  0.075}_{-  0.100}$ & $  2.484^{+  1.260}_{-  1.500}$ & ... \\
 1 & 0.02 -- 0.10 &  -1.664 & -21.683 & ... &  -3.158 \\
 2 & 0.10 -- 0.25 & ... & ... & ... &  $   -2.924^{+   0.025}_{-   0.057}$\\
 3 & 0.25 -- 0.40 & ... & ... & ... &  $   -3.144^{+   0.029}_{-   0.064}$\\
 4 & 0.40 -- 0.60 & ... & ... & ... &  $   -3.078^{+   0.017}_{-   0.061}$\\
 5 & 0.60 -- 0.80 & ... & ... & ... &  $   -3.029^{+   0.013}_{-   0.064}$\\
 6 & 0.80 -- 1.20 & ... & ... & ... &  $   -3.123^{+   0.018}_{-   0.067}$\\
 7 & 1.20 -- 1.50 &  -1.664 & -22.749 & ... &  $   -3.688^{+   0.066}_{-   0.085}$\\
 8 & 1.50 -- 2.50 &  -1.664 & -22.967 & ... &  $   -2.769^{+   0.008}_{-   0.006}$\\
 9 & 2.50 -- 3.50 &  -1.664 & -22.967 & ... &  -3.252 \\
10 & 3.50 -- 6.00 &  -1.664 & -22.967 & ... &  -3.280 \\
\hline
\end{tabular} \\
Results from the fit of SED-specific Schechter
  functions to the redshift range $0.1 \leq z < 1.2$. In the first row of each section of the table, we give the shape parameters
  describing the best-fit LF in this range. Normalization parameters
  $\Phi^{*}$ are listed separately for each redshift bin. Values
  without uncertainties are constant background values and not
  optimized. The parameter M$^*$ is evaluated at $z=0.95$.
\label{tab_cosmiclf}
\end{minipage}
\end{table*}

\subsection{The LF in individual redshift bins}
\label{sec_individualzbins}
Our best-fit solution described in \S \ref{sec_cosmiclf} has been based on a relatively low
number of free parameters. One of the disadvantages incurred by this
is that the constraints on $M^{*}$ are weak, and the extrapolation to redshift
zero via the two parameters $M^{*}_0$ and $\delta$ is potentially
affected by large errors. Furthermore, our assumptions (such as the
simple two-parametric model for $M^{*}$ as a function of redshift) cannot be verified
explicitly within the confines of this simple approach.

To explicitly test these assumptions and obtain more accurate
constraints on the LF at low redshift, we now refine our solution by maximizing the likelihood with
respect to the parameters $M^{*}$ and $\Phi^{*}$ in individual
redshift bins. We iterate both parameters for each SED type and for
each redshift bin. Other than that, the same parameters are included
in this optimization as in the previous section: $\Phi^{*}$ in
redshift bins 7 and 8, and the faint-end slope $\alpha$ in bins 2 to 6
(assumed independent of redshift).

\begin{figure}
\includegraphics[width=84mm]{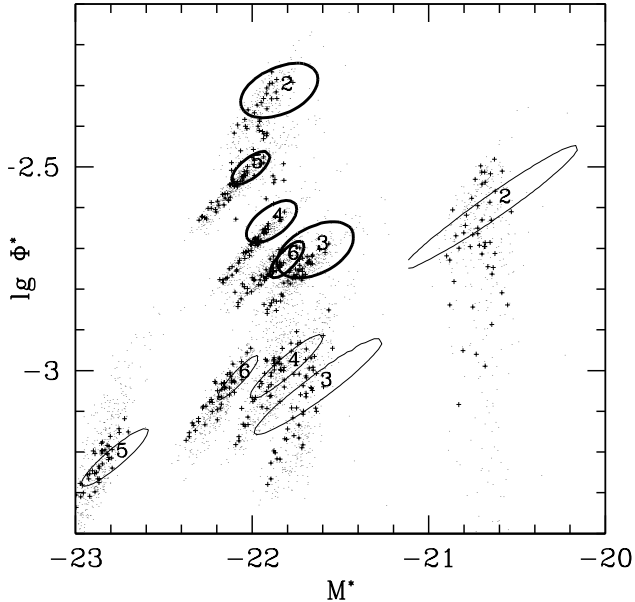}
\caption{Uncertainty contours at the 68\% confidence level
  for fits to the early-type (bold contours) and late-type (light contours) LF in individual redshift bins over the range $0.1 \leq z < 1.2$. For the early-type LF, we see no evidence
  for an evolution of $M^{*}$, while $\Phi^{*}$ exhibits considerable
  scatter, which may be associated with known density inhomogeneities
  in the E-CDFS. The late-type LFs exhibit a strong covariance between
$M^*$ and $\Phi^*$ and populate a sequence ranging from faint
  $M^*$/high $\Phi^*$ at low to bright $M^*$/low $\Phi^*$ at high
  redshift, attesting to evolution in at least one parameter. Dots and
crosses illustrate how variations in the constant high- and
low-redshift LF parameters may propagate into the recovered LFs.}
\label{fig_iphims}
\end{figure}

The results of this analysis --- in an $M^{*}$-$\Phi^{*}$ projection
--- are shown in Fig. \ref{fig_iphims}. Bold contours represent early
types, and thin contours late types. Error contours are calculated for
two degrees of freedom and 68\% confidence. Table \ref{tab_refinedlf} gives the best-fit values in
individual redshift bins.

The early-type
galaxies exhibit relatively little scatter in $M^{*}$ --- all error
contours lie in the range $-22.2 < M^{*}_{r} < -21.5$. Given the
relatively large uncertainties, there is no obvious trend of $M^{*}$ with redshift, confirming the conclusion that $\delta\approx0$. The scatter in $\Phi^{*}$ is larger than the
uncertainties; in the seven redshift bins included in our analysis,
$\Phi^{*}$ scatters within a factor of $\sim 3$. The scatter may
be affected by large scale structure. \citet{phleps} report an
underdensity at $0.25 < z < 0.4$, which coincides with our very low
data point in redshift bin 3. On the other hand, \citet{adami05},
based on spectroscopic data, note several structures, including
massive galaxy groups, at $z\approx 0.7$, which falls into our redshift bin
5, which is significantly elevated in comparison to the neighbouring
bins. Our LF results are therefore qualitatively consistent with 
large scale structure features in the E-CDFS known from spectroscopic studies. The scatter introduced
by these features is too large to permit any reliable conclusion as to
whether $\Phi^*$ exhibits a trend with redshift. 

The uncertainty contours for the late-type LFs are affected by the
strong covariance between $M^{*}$ and $\Phi^*$ in the case of a steep
faint end. The best-fit values form an approximate sequence, with the faintest, but highest-normalization
fits being found at low redshift, while the high-redshift LFs are
brighter by more than two magnitudes in $M^{*}$, but also exhibit a
lower $\Phi^{*}$ (by a factor of $2$---$3$); this is also consistent with
our earlier finding of a positive evolution parameter
$\delta$. Comparison of the corresponding Schechter profiles shows
that the variation of the late type LF between redshift bins occurs
primarily at the bright end, whereas the normalization of the LFs in
this redshift range is almost constant with redshift at magnitudes
$M_r>-20$. However, redshift bin 3 deviates significantly towards lower
normalizations from the sequence along which the uncertainty contours
for the other bins are aligned, which is again consistent with the
claim of an underdensity in the E-CDFS in this redshift range.

We have again determined the impact of varying the high- and
low-redshift LF parameters held constant during these fits using a
Monte Carlo procedure as described in \S \ref{sec_errors}. The range of
best-fit parameters is shown in Fig. \ref{fig_iphims} by dots for all
trials, and crosses for trials yielding acceptable likelihood. For the
early-type LFs, shifts towards slightly brighter $M^*$/lower $\Phi^*$
are possible, but shifts in the opposite sense (fainter $M^*$) are
ruled out. For the late-type sample, the background uncertainties
propagate primarily into variations in $\Phi^*$. We remind the reader
that the magnitude of these variations is dependent on the magnitude
of the high- and low-z background uncertainties, which are assumed to be
very generous here (a factor of 10 in $\Phi^*$, and one mag in
$M^{*}$), and that the Monte Carlo procedure can therefore only convey
a qualitative impression of the robustness of the results and the
sense in which they may deviate from the default solution.

\begin{table}
\caption{LF Results at $0.1 \leq z < 1.2$}
  \begin{tabular}{@{}llrrrr}
  \hline
bin & z & $\alpha$ & M$^{*}$ & lg $\Phi^{*}$ \\
\hline
 &early types \\
 1 & 0.02 --- 0.10 & $-0.639$ & $-21.92$ & $ -2.502$ \\
 2 & 0.10 --- 0.25 & $-0.640$ & $-21.84^{+  0.20}_{-  0.20}$ & $ -2.311^{+  0.063}_{-  0.063}$\\
 3 & 0.25 --- 0.40 & $-0.640$ & $-21.64^{+  0.20}_{-  0.20}$ & $ -2.702^{+  0.063}_{-  0.063}$ \\
 4 & 0.40 --- 0.60 & $-0.640$ & $-21.89^{+  0.13}_{-  0.13}$ & $ -2.633^{+  0.038}_{-  0.050}$ \\
 5 & 0.60 --- 0.80 & $-0.640$ & $-22.01^{+  0.10}_{-  0.10}$ & $ -2.502^{+  0.038}_{-  0.038}$\\
 6 & 0.80 --- 1.20 & $-0.640$ & $-21.80^{+  0.07}_{-  0.07}$ & $ -2.727^{+  0.038}_{-  0.037}$ \\
 7 & 1.20 --- 1.50 & $-0.639$ & $-21.87$ & $ -2.721^{+  0.051}_{-  0.055}$ \\
 8 & 1.50 --- 2.50 & $-0.639$ & $-21.85$ & $ -2.694^{+  0.071}_{-  0.079}$ \\
 9 & 2.50 --- 3.50 & $-0.639$ & $-21.83$ & $ -2.850$ \\
10 & 3.50 --- 6.00 & $-0.639$ & $-21.80$ & $ -2.953$ \\
\hline
 &late types\\
 1 & 0.02 --- 0.10 & $-1.664$ & $-21.68$ & $ -3.158$ \\
 2 & 0.10 --- 0.25 & $-1.637$ & $-20.60^{+  0.43}_{-  0.50}$ & $ -2.588^{+  0.138}_{-  0.150}$ \\
 3 & 0.25 --- 0.40 & $-1.637$ & $-21.62^{+  0.32}_{-  0.35}$ & $ -3.037^{+  0.100}_{-  0.112}$ \\
 4 & 0.40 --- 0.60 & $-1.637$ & $-21.80^{+  0.18}_{-  0.20}$ & $ -2.987^{+  0.063}_{-  0.075}$ \\
 5 & 0.60 --- 0.80 & $-1.637$ & $-22.77^{+  0.17}_{-  0.18}$ & $ -3.212^{+  0.063}_{-  0.063}$ \\
 6 & 0.80 --- 1.20 & $-1.637$ & $-22.08^{+  0.10}_{-  0.10}$ & $ -3.017^{+  0.050}_{-  0.050}$ \\
 7 & 1.20 --- 1.50 & $-1.664$ & $-22.75$ & $ -3.633^{+  0.070}_{-  0.081}$ \\
 8 & 1.50 --- 2.50 & $-1.664$ & $-22.97$ & $ -2.700^{+  0.006}_{-  0.007}$ \\
 9 & 2.50 --- 3.50 & $-1.664$ & $-22.97$ & $ -3.252$ \\
10 & 3.50 --- 6.00 & $-1.664$ & $-22.97$ & $ -3.280$ \\
\end{tabular}\\
Best-fit LF parameters in individual redshift
  bins. Fits are carried out in bins 2--8. Best-fit values for
  $\Phi^{*}$ in redshift bins 7 and 8 are shown for completeness, but
  lie beyond the reliable redshift region. 
\label{tab_refinedlf}
\end{table}

\subsection{Comparison to other surveys}
\label{sec_comparison}

Our results for the low-redshift LF appear to be in contrast to
other determinations for the local field LF \citep{blanton2003}; in particular,
$M^{*}$ for the early-type population at low redshift appears strikingly bright,
compared to a value of $M^{*}_r=-21.18\pm0.01$ at z=0.1 given by Blanton et
al. (adjusted to $H_0=71$). 

\begin{figure}
\includegraphics[width=84mm]{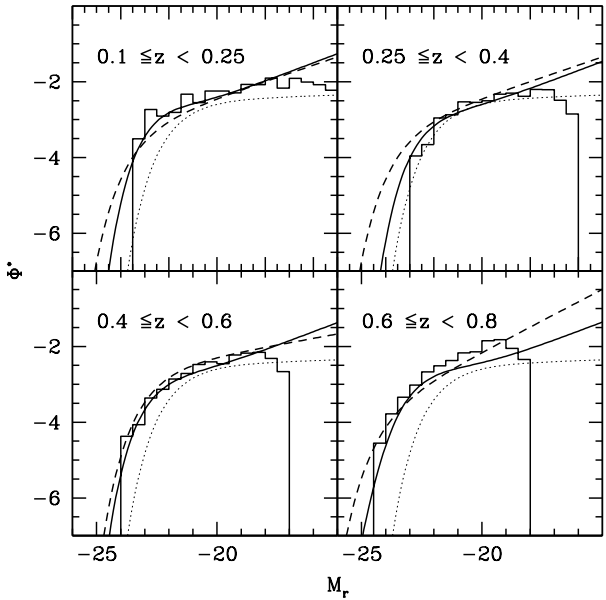}
\caption{Comparison between the distribution of spectroscopically
  confirmed galaxies in absolute magnitude (histogram) with our LF fits to the
  photometric catalog (solid line), the z=0.1 field LF from
  \citet{blanton2003} (dotted line), and LFs in similar redshift bins from \citet{wolf} (dashed line). At the bright end, it confirms the shape
  of the PML solution. At the faint end, the observed distribution is
  a lower limit, because no volume correction has been applied.
}
\label{fig_blantontest}
\end{figure}

To illuminate the causes of this discrepancy, we present in
Fig. \ref{fig_blantontest} a comparison between the Blanton et al. LF
at z=0.1 and our results in four low redshift bins, as given in Table
\ref{tab_refinedlf}. For the purpose of facilitating comparison, we
have drawn the $z=0.1$ Blanton et al. LF in all four panels, but emphasize that even the
lowest-redshift bin of our study shown here lies above the redshift
for which that LF was calculated. 

The histograms show the distribution of absolute magnitudes of
galaxies with known spectroscopic redshifts in the MUSYC catalog. For
bright galaxies, more than 25\% of the objects in the catalogue have
spectroscopic redshifts; we have applied an approximate completeness
correction as a function of the apparent magnitude $m_R$ and divided
by the comoving volume of the redshift bin. Apart from the calculation
of the rest-frame $r$-band absolute magnitudes, these are the only
corrections applied to the data. In particular, we have abstained from
applying a volume correction in order to minimize the risk of introducing systematic errors into
this distribution. For this reason, the
distribution of observed galaxies can only give a reasonable
impression of the shape of the LF at the bright end, which is observable across the entire bin.

The observed distribution does not agree with our LFs at the faint
end, which is not surprising, because no volume correction has been
applied; the observed distribution is therefore
likely to be a lower bound on
the true LF, because the prevalence of faint galaxies, which are not
visible across the entire volume of a given redshift bin, has
certainly been underestimated.

Nonetheless, Fig. \ref{fig_blantontest} reveals a remarkably good
agreement between our best-fit LFs calculated with the PML and the directly observed galaxy distribution, both
with regard to the shape of the bright end and the normalization. In
comparison, in the lowest-redshift bin, which contains the redshift of
$z=0.1$ for which the Blanton et al. LF holds, the observed
distribution of absolute magnitudes extends to far brighter magnitudes
than that LF would permit. 

The disagreement with the low-redshift LF of Blanton et al. is
surprising. It cannot be due to a specific fault of our LF algorithm,
because even just the absolute magnitude distribution of
spectroscopically confirmed galaxies is clearly in excess of the
former's prediction. The fault does not appear to lie with our
calculation of absolute magnitudes either, because comparison to
absolute magnitudes published in the COMBO-17 catalog shows at most
small ($\Delta M \approx -0.06$) systematics even as far as
$z\approx1$ and scatter in the recovered absolute magnitudes of $\sim
0.22$, not nearly enough to push the bright end forward by about a
magnitude. At the extreme low-redshift end, our recovered absolute
magnitudes are also in excellent agreement with comparison values
calculated naively from the observer-frame $r$-band magnitudes simply
by applying a distance modulus, indicating that our procedure of
calculating absolute magnitudes from SED fits to all filter bands
simultaneously does not introduce discrepancies from the conventional
procedure of applying a $k$-correction to the measured photometry in a
single band.

In this context, it is worth pointing out that the LFs recovered by \citet{wolf} from COMBO-17 (which was obtained with
the same telescope and instrument and overlaps with the E-CDFS field that we analyze here) are also significantly brighter than the Blanton et
al. estimate and as bright as or even brighter than our solution. For comparison, we have overplotted the total LFs of
Wolf et al. in Fig. \ref{fig_blantontest} as dashed lines. In the
upper left and upper right panels (covering the redshift ranges $0.1
\leq z < 0.25$ and $0.25 \leq z < 0.4$, respectively, the Wolf et
al. LFs shown are those at $z=0.3$ (the lowest redshift bin
independently constrained by their work), in the bottom left panel, it
is the LF at $z=0.5$, and in the bottom right panel, $z=0.7$, which
coincides well with our choice of binning. In all bins, the Wolf et
al. LF is substantially brighter than the estimate of Blanton et
al. at $z=0.1$, and agrees qualitatively with or is even brighter than our LF determination. The faint-end slopes in their work show considerable variance, but in the lowest-redshift bin, where it may be supposed to be constrained best, it matches ours in normalization and slope.

The agreement between our work and that of \citet{wolf} with regard to
the bright end at low redshift being substantially brighter than
estimated by \citet{blanton2003}, suggests that the depth of the
photometry, and particularly the surface brightness depth, may be at
least partly responsible for the brighter faint end recovered by us
and \citet{wolf}. In particular, it is possible that the often very
extended envelopes of luminous early-type galaxies, which dominate the
bright tip of the LF, might be underestimated in the shallower SDSS
data.

Sample variance due to large scale structure must also be taken into
account as a possible additional contributor to the discrepancy
between our LF and that of Blanton et al.; Fig. \ref{fig_blantontest}
attests to considerable variance from one redshift bin to the
next. We do recover a significantly fainter LF in the redshift range
$0.25 \leq z < 0.4$ than in the $0.1 \leq z < 0.25$ bin; however, this
bin appears to be exceptionally faint and underdense.

Variations in the high- and low-z background parameters that have so
far been held constant will not shift $M^*$ to significantly fainter
values, as we have argued in \S \ref{sec_cosmiclf}.

\subsection{The evolution of the luminosity density}
\label{sec_lumden}
Covariances between many of the parameters involved in describing
the shape and normalization of the LF --- such as between $M^{*}$ and
$\delta$, or $M^{*}$ and $\Phi^{*}$ --- imply that a fairly large range
of parameters can reproduce LFs with essentially similar shapes and
evolutionary behaviour. For example, for a very steep faint end slope
$\alpha$, $M^{*}$ and $\Phi^{*}$ are highly degenerate, so that the
same evolution with $z$ can be described either as an evolution of
$M^{*}$ or of $\Phi^{*}$.

\begin{figure}
\includegraphics[width=84mm]{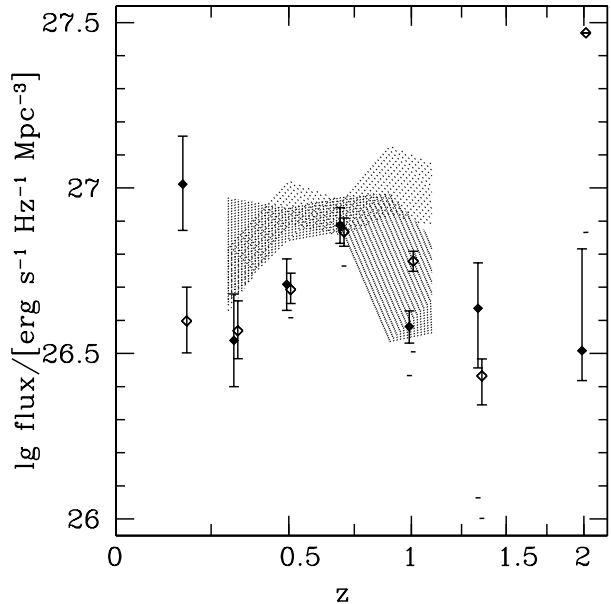}
\caption{Luminosity density as a function of redshift for early types
  (filled diamonds) and late types (empty diamonds). Shaded regions
  show values from \citet{wolf} for comparison (light for late, dark
  for early types). Data points above z=1.2 are included in the
  optimization, but lie beyond our pre-defined redshift range of
  confidence. Horizontal bars show the luminosity density integrated
  approximately to the effective magnitude limit.
}
\label{fig_lumden}
\end{figure}

Specifically for the comparison to other surveys and to galaxy
evolution models, it is thus desirable
to transform the derived LF parameters to another quantity that does
not suffer from these degeneracies. One such quantity is the
luminosity density, which can be obtained by integrating the
luminosity-weighted LF. We do so in each redshift bin and for early
and late type objects separately. We integrate the LFs from Table \ref{tab_refinedlf} to a limit of $M_R=-14$. Assuming a reasonable
extrapolation of the LF to fainter magnitudes, this limit includes
almost the total luminosity density, even for the steep LF of late
type galaxies. The results are shown in Fig. \ref{fig_lumden}. Empty
data points illustrate the luminosity density of late-type galaxies,
and filled points of early-type galaxies. Error bars are calculated
from the range of models within the 1$\sigma$ uncertainty contours in the $M^{*}$-$\Phi^{*}$
plane. Covariances with $\alpha$ are neglected, but the luminosity
density is typically dominated by galaxies near $M^{*}$, so that the
actual contribution of the faint end to the luminosity density is very
minor. To illustrate this, horizontal bars plotted above $z=0.4$ show
the luminosity density in each bin integrated only to the approximate effective absolute magnitude limit
(determined from Fig. \ref{fig_maglimits}). For the early type objects,
the difference is minute up to $z=1$. For late types, the difference
is more appreciable because of the much steeper faint end; above
$z=1$, at most half of the projected luminosity density would be contributed by galaxies within our survey limits. Nonetheless, our qualitative conclusions are not affected by this.

As we noted previously in \S \ref{sec_individualzbins} with regard to the
evolution of $\Phi^*$, the presence of density inhomogeneities in the
field adds too much scatter to quantify how the luminosity density
evolves with redshift. We will therefore address this question again in a
subsequent analysis based on the full MUSYC sample, drawn from four fields in
different regions of the sky. At $z>1.2$, large jumps in the luminosity density
of late types in redshift bins 7 and 8 reflect the problems that we
have alluded to in \S \ref{sec_cosmiclf}, namely, that these bins fall
outside the trusted redshift region defined in that section, and
are only included in the optimization due to the strong covariances
between neighbouring bins, in order not to restrict the number of
degrees of freedom in the search for the maximum likelihood solution
unreasonably, while their best-fit values themselves are
questionable. In particular, the very high value for the luminosity
density of late-type galaxies in redshift bin 8 is likely to be an
artifact.

 For comparison, we have indicated the $r$-band
luminosity densities of \citet{wolf} (transformed to $H_0=71$ and after
reclassifying their Types 1 and 2 as early types and Types 3 and 4 as
late types) as dark (early types) and light (late types) shaded areas. On average,
their luminosity densities are higher than ours by $\sim 0.2$ dex. The results
agree very well in our redshift bin 5, which comprises one of the
known large scale structure features, but their data do not show the
strong underdensity in the region $0.25 \leq z < 0.4$ as pronouncedly
as ours do.

\subsection{The detailed shape of the LF}
\label{sec_binnedlf}

A striking feature of our LF solution is the extremely steep faint end
of the LF of late-type galaxies of $\alpha\approx-1.62$. This steep
slope dominates the total LF at the faint end, leading to an upturn at
intermediate magnitudes. However, even this double-Schechter shape may not provide a
comprehensive description of the faint end of the LF, because even within the
subclasses of early and late types that we analyze, features that are
not consistent with the Schechter functions, such as up- or downturns,
may exist. In particular, galaxy evolution models \citep{springel2001,
  delucia2004}, which propose a host of mechanisms for suppressing star formation in low mass halos, and the
relatively low number counts of ultra-faint dwarf galaxies in the
Local Group (SDSS), which show good agreement with such models
\citep{maccio,koposov}, may require a break in the late-type LF if they
are to be brought into consistency with our claim of a very steep
faint end slope at intermediate magnitudes. This raises two questions:
Down to what absolute magnitude can our analysis constrain the LF, and is our best-fit Schechter function
really an adequate fit over this entire range, or are there
indications of a departure from this shape at faint luminosities?

This question is answered best by representing our LF by a binned
distribution, rather than by a low-parameter analytical {\it
  ansatz}. As in \S \ref{sec_cosmiclf}, we solve for the LF over the
range $0.1 \leq z < 1.2$, but instead of modeling the SED-specific LF
with evolving Schechter functions, we describe it with series of bins
in absolute magnitude of 1 mag width each. The bin boundaries are taken relative to $M^*$
in a given redshift bin, and the normalization of the LF value $\Phi$
in each magnitude bin is also relative to the normalization parameter
$\Phi^{*}$ in a redshift bin, as inferred from our solution in Table
\ref{tab_cosmiclf}. Therefore, the binned solution represents the
shape of the LF after removing the evolution in $M^{*}$ and
$\Phi^{*}$. All LF parameters not explicitly represented by this
binned ansatz are held constant at their best-fit values from Table \ref{tab_cosmiclf}.

For our first analysis, we scan the likelihood distribution for one
magnitude bin at a time as a function of $\Phi$ (the value of the LF
across this bin), while keeping all other bins at the values implied by the Schechter function. The
resulting uncertainty range is a measure of how much an individual
magnitude bin may deviate from the Schechter fit, and will thus
provide very conservative upper and lower limits on the shape of the
LF at the faint end. If these limits are sufficiently tight, they will
allow us to rule out major departures from the Schechter function.

\begin{figure}
\includegraphics[width=84mm]{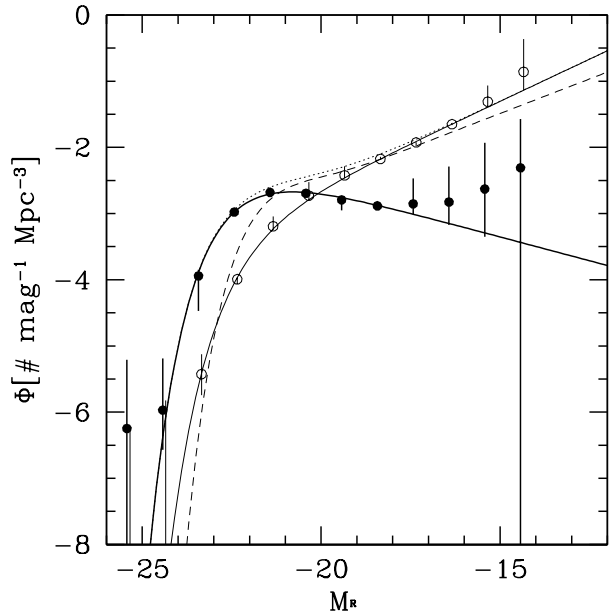}
\caption{Binned representation of the shape of the late- and
  early-type LFs at $0.1 \leq z < 1.2$, shifted to $M^{*}$ and
  $\phi^{*}$ in redshift bin 2. Magnitude bins have been optimized one
  at a time. The thick solid curve represents the
  best-fit Schechter function of early types, the thin curve of late
  types. Data points with error bars show the uncertainties on $\Phi$
  in any one individual magnitude bin under the (idealized) assumption
  that all bins are independent. They serve as an upper limit on any
  deviation of the LF shape from the Schechter function, and show that
  the late-type LF does not exhibit a break down to at least
  $M_{r}\approx-14$, but instead continues to rise steeply with
  $\alpha\geq-1.65$. The dotted line represents the sum of both LFs,
  and the dashed line the steepest LF suggested by
  \citet{blanton2005}, which agrees well with ours around the onset of
  the faint end upturn, although it fails to match our LF at the
  bright end.
}
\label{fig_cosmicsedlf}
\end{figure}

Fig. \ref{fig_cosmicsedlf} shows the results of this calculation. The
LF has been adjusted to the $M^*$ and $\Phi^*$ values in redshift bin
2 from Table \ref{tab_cosmiclf}. Thin and thick solid lines show the
best-fit Schechter functions, as calculated previously, for late and
early types, respectively, and the dotted line illustrates their sum,
which is dominated by early times at the bright end and by late types
at the faint end. Data points (filled for early types, empty for late
types) show the maximum likelihood value for each bin, and error bars
the acceptable $68\%$ confidence range for one degree of freedom. 

The superposition of the shallow early and steep late type LFs
produces a distinct upturn at intermediate magnitudes of $M_r\approx-19$.
Such an upturn has been suggested by \citet{blanton2005} on the basis
of SDSS data. In the latter survey, the fairly shallow surface
brightness limit prevents a direct measurement of the full faint end
of the LF, but extrapolating for galaxies beyond this limit, their
steepest estimate for the faint-end component is $\alpha=-1.52$,
slightly shallower than our result. This estimate is shown in
Fig. \ref{fig_cosmicsedlf} by a dashed line. Around the onset of the
faint-end component, at $M_r\approx-19$, our LF agrees very well in
normalization and shape with the LF of Blanton et al. The Blanton LF does deviate significantly from ours at
the bright end, a phenomenon already discussed in \S \ref{sec_comparison}.

The principal conclusion from this figure is that we can constrain the
late-type LF to $M_{r}\approx-14$, and the early-type LF to
$M_{r}\approx-15$. A downturn in either LF above these magnitude limits
can be ruled out. If mechanisms suppressing star formation do act to
lower the abundance of dwarf galaxies from the extrapolated $\alpha\approx-1.62$
LF, this must happen at even fainter absolute magnitudes. There is some room
for an excess of faint galaxies over the Schechter fits.

\begin{figure}
\includegraphics[width=84mm]{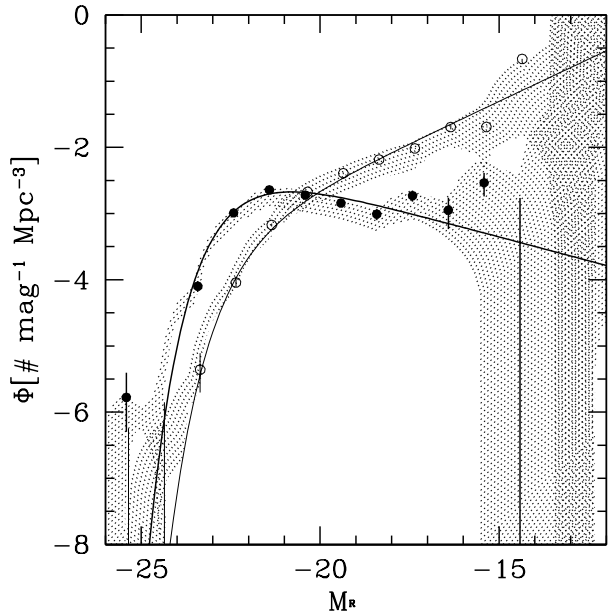}
\caption{Binned representation of the shape of the late- and
  early-type LFs at $0.1 \leq z < 1.2$, shifted to $M^{*}$ and
  $\Phi^{*}$ in redshift bin 2. Here, all magnitude bins have been
  optimized simultaneously, representing our best estimate for the
  shape of the LF with associated uncertainties around the best
  fit. Shaded areas show how variations in the high- and low-redshift
  background LF parameters propagate into the best-fit solution.
}
\label{fig_cosmicsedlfit}
\end{figure}

To gain a more accurate picture of the shape of the LF, we now carry
out a simultaneous optimization of all magnitude bins of both
SEDs. After finding the best-fit solution, we obtain uncertainties
with the previously-described strategy (for each free parameter, we
scan the range of values that yield an acceptable offset in
$\chi^{2}$ from the best-fit solution, allowing all other magnitude bins
to be optimized in the background). The resulting LF is shown in
Fig. \ref{fig_cosmicsedlfit}. Error bars are again calculated for one
degree of freedom and a confidence level of 68\%, as appropriate for an
error calculation in a single parameter. The uncertainties are
generally smaller than in the previous analysis, which at first
appears counterintuitive (allowing the ``uninteresting'' parameters to
find a conditional maximum likelihood generally enlarges the
uncertainties on any one ``interesting'' parameter). However, this can
be explained by the fact that we are now evaluating the uncertainties
relative to the true maximum likelihood fit in the parameter space
consisting of the individual magnitude bins, rather than the best Schechter
fit.

This binned LF is sensitive to our assumptions for the
values of the constant ``background'' LF parameters at very low and
very high redshift. To illustrate this sensitivity, we have taken the
output from the Monte Carlo trials discussed in \ref{sec_cosmiclf} and
calculated the binned LF for those trials not rejected due to their low
likelihood. The shaded areas in Fig. \ref{fig_cosmicsedlfit} show the
full range covered by these binned LFs. 

The figure confirms the basic picture from
Fig. \ref{fig_cosmicsedlf}: The late-type LF continues to rise steeply
to $M_{r}\approx -14$ without any indication of a break. Of all our Monte
Carlo trial runs with acceptable likelihoods, not one exhibits a
strong downturn in the faint end slope at $M_{r}<-14$. Although the
last two significant bins in the late-type LF do exhibit small, but
significant deviations from the Schechter fit, the Monte Carlo results
show that variations in the high- and low-redshift background can
account for their magnitude. 

The early-type LF is constrained well down to $M_{r}=-16$, and upper
limits can be applied down to $M_{r}=-14$. These show that the LF follows
our Schechter fit to at least $M_{r}=-16$. The uncertainties leave room for
deviations from the power-law faint end slope in either sense at
fainter magnitudes, but nonetheless the late-type
LF clearly dominates the total LF at least to $M_r<-14$. 

We have repeated this analysis using the more stringent
star/galaxy classifier limit of 0.8, instead of our default 0.97. This
incurs the cost of possibly excluding very compact dwarf galaxies, but
is closer to typical values adopted in the literature. As we have seen
in \S \ref{sec_cosmiclf}, the best-fit early-type LF is characterized by a
slightly shallower faint-end slope and fainter $M^{*}$, while the
late-type LF is almost unaffected. However, calculation of the binned
LF in the manner described above shows no qualitative difference in
the shape of the LF, compared to our default calculation, which indicates that the
recovered LF shape does not depend on the precise star/galaxy
threshold (at least in the range $0.8 \leq S/G \leq 0.97$, which is
spanned by these two samples).

\section{Conclusions}

We have proposed a new Bayesian method for the calculation of galaxy luminosity
functions from multi-band photometric samples. The method, the {\it
  Photometric Maximum Likelihood Method}, deconvolves the observed
distribution of galaxies as a function of fluxes in multiple filter
bands into the luminosity functions of all galaxy populations
(identified by redshift and SED type) that may contribute to this
observed distribution. It thus eliminates the requirement to assign
photometric redshifts to individual galaxies. This approach offers
multiple advantages:

\begin{itemize}
\item The PML achieves a proper deconvolution of the observed data
  into the constituent galaxy populations.
\item The entire information contained in the multi-band photometric
  data is retained for the calculation of the LF, instead of incurring
  the information loss that is associated with reducing these
  photometric data to a single best-fit photometric redshift.
\item There is no requirement to model the often very complicated
  error functions of photometric redshifts; instead, only an
  understanding of the photometric error function is required.
\item Conventional photo-z determinations and LF calculations based on
  them are dependent on many (implicit or explicit) priors about the
  galaxy distribution as a function of redshift, which often go unacknowledged. By solving for the
  parameters of multiple galaxy populations over a large range of
  redshifts simultaneously, the PML significantly reduces the number
  of priors and allows for a self-consistent solution.
\end{itemize}

We have applied this algorithm to the E-CDFS field of MUSYC (the
Multi-Wavelength Survey Yale-Chile) with the aim of recovering the
faint-end shape of the LF at low redshift. For this purpose, we have
represented the galaxy population up to $z=1.2$ by two SED-specific
Schechter functions (one for early-, one for late-type
galaxies). Principal results are:

\begin{itemize}
\item In accordance with the current consensus in the literature, the
  LF of early-type galaxies is characterized by a fairly bright
  characteristic absolute magnitude $M^{*}_{r}$ and a shallow faint
  end slope, while the LF of late-type galaxies
  is much steeper.
\item The faint-end slope of the late-type galaxy population may be as
  steep as $\alpha\approx -1.62$. The superposition of these two
  SED-specific LFs produces an overall LF that cannot be described by
  a single Schechter function, but instead exhibits an upturn at
  intermediate magnitudes, where the late-type LF begins to dominate
  the overall LF. This upturn is noticeable in comparison to a
  standard $\alpha=-1.2$ LF at magnitudes fainter than $M\approx
  M^{*}+3$ ($M\approx -19$ in our sample).
\item The bright-end exponential cutoff is significantly
  brighter than found in the SDSS \citep{blanton2003}, but
  slightly fainter than found in COMBO-17 \citep{wolf} from photometric
  redshifts. 
\item We find fluctuations in the normalization of the LF and in the
  luminosity density with redshift in the E-CDFS that are broadly
  consistent with large-scale density inhomogeneities known from
  spectroscopic data but prohibit firm conclusions regarding any
  systematic trend of these quantities with redshift.
\item Calculation of the shape of the LF using a binned {\it ansatz},
  rather than the analytical Schechter form, shows that the steep
  faint end extends at least as faint as $M_{r}\approx-14$, which
  confirms the presence of a faint end upturn. Therefore,
  we postulate a significantly larger number of dwarf galaxies in the universe than
  would be predicted by simply extrapolating the typical faint end
  slope of $\alpha\approx-1.2$ recovered in shallower field galaxy
  surveys to fainter magnitudes.
\item  Our results show that,
  although our conclusions are qualitatively robust, statistical results
  still depend on priors, such as the extreme high-redshift LF, that are difficult to constrain in
  a self-consistent way, even with a high-quality data set such as
  MUSYC. Any work attempting to constrain these quantities from purely
  photometric data must give a careful account of the effect of such
  priors on its conclusions. 
\end{itemize}

The PML formalism offers many opportunities for refinements. It can be
easily generalized to calculate distribution functions over colors,
stellar masses, or star formation rates, as long as these quantities
can be unambiguously assigned to the SED templates used in the
analysis. To improve the calculation of LFs at high redshift, it is
possible to implement passive redshift evolution in the SED templates,
as opposed to the redshift-zero empirical templates which we have used
in our analysis. We are planning a subsequent study of the full MUSYC
catalog which will explore some of these avenues.

\section*{Acknowledgments}

DC gratefully acknowledges financial support from the {\it Fundaci\'on
  Andes} during the initial work on this project. DM is supported by
NASA LTSA NNG04GE12G. EG is supported by the National Science
Foundation under grant AST-0807570. We also acknowledge stimulating discussions with
Edward Taylor and a helpful report from an anonymous referee that has
greatly improved this manuscript.

\appendix
\section{Modifications to the published MUSYC photometric catalog}
\label{sec_photcorr}

We have made two modifications to the published catalogs of the MUSYC-ECDFS in order to
optimize the accuracy of the photometric redshifts, and, by extension,
of our LF results. The first of these is a {\it zero-point
  correction}. Zero-point offsets manifest themselves in photometric
fluxes in one given filter band generally being determined higher or
lower than the true fluxes by a constant factor, and can be due to an
inaccuracy in the photometric calibration or the transmission
curves. In order to detect such offsets, we have carried out SED fits
to all galaxies with known spectroscopic redshifts. The residuals of
the observed fluxes relative to the fluxes predicted from the best-fit
SED are an approximation of the zero-point offset of the fluxes
recorded through this filter. By visual inspection of the distribution
of photometric residuals, we have determined the required corrections
to the published fluxes as a function of the filter band and list them
in Table \ref{tab_photcorr}. The fluxes are corrected according to the
formula
\begin{equation}
f_{corr} = f_{published} * 10^{-0.4 \times \Delta} .
\label{eq_fluxcorr}
\end{equation}

This procedure is also sensitive to template insufficiency, i.e., an
inability of the template set to reproduce the entire range of SEDs
displayed by our sample objects. Ascribing systematic photometric residuals that are really caused by template insufficiencies to photometric zero point shifts instead may optimize the quality of photo-z fits to the spectroscopic subsample, but possibly introduce systematic errors for other galaxy populations and other redshifts. In order to provide an independent verification of our zero-point shifts, we have applied a second test that is less sensitive to template insufficiencies: We have compared photometric residuals between two different filters, but for the same SED and rest-frame wavelength, i.e., we have examined the distribution of the quantity
\begin{eqnarray}
\delta \Delta = lg\:\frac{f_{obs}}{f_{pred}} \mid_{1} - lg\:\frac{f_{obs}}{f_{pred}} \mid_{2} = lg\:\Delta(filter_1) - lg\:\Delta(filter_2) \nonumber \\
\qquad  + lg\: c_2(SED,\lambda_{rest})\mid_1-lg\: c_2(SED,\lambda_{rest})\mid_2
\end{eqnarray}

for all possible pairings of galaxies fitted with the same SED template and as a function of the difference in rest-frame wavelength. The term $c_2$ is a hypothetical correction to the SED template, dependent on the template itself and the rest-frame wavelength. If the difference in $\lambda_{rest}$ between the two observations is small and both are drawn from the same SED, the third and fourth logarithms cancel, and this becomes a measure of the difference in photometric residuals, evaluated for the same SED type at the same $\lambda_{rest}$, but evaluated through different filters at different redshifts. In practice, the corrections $c_2$ will depend on the true SED of the observed objects, which may deviate from the best-fit SED template, so that the logarithms that depend on $c_2$ do not cancel perfectly, but the size of such deviations is limited by the fact that both objects are fitted by the same SED template, so that this method allows for at least a qualitative verification of the previously found zero-point shifts.

The resulting relative zero-point shifts between any two optical filter bands are consistent with the numbers presented in Table \ref{tab_photcorr}, bolstering our confidence that the photometric residuals can indeed be predominantly ascribed to zero-point adjustments, rather than template insufficiencies. We do point out that we have not been able to use this method to independently verify our zero-point shifts for the $J$- and $K$-bands, because the number of objects permitting us to observe the same $\lambda_{rest}$ through both filters at different redshifts is too small. However, the actual impact of these zero-point corrections on our LF results is minor and does not alter our qualitative conclusions.

The second adjustment to the photometric catalogue concerns the catalog uncertainties. We have tested the catalog MUSYC photometric uncertainties by comparing the scatter of the residual photometric fluxes from SED fitting to objects with known redshifts. While, in the ideal case, the residuals should show a Gaussian distribution (i.e., 68\% should fall within the published standard deviation, 95\% should fall within two standard deviations, etc.), the residuals in the MUSYC catalog typically exhibit a slightly larger scatter, indicating that the catalog uncertainties are an underestimate, and the scatter increases with increasing flux. The latter behaviour is easily understood, because the published uncertainties only reflect internal photometric consistency, but do not include photometric calibration uncertainties or local background errors.

In order to achieve a more realistic, Gaussian distribution of the uncertainties, we have introduced two additional parameters to transform the published errors:
\begin{equation}
\sigma_{new} = \sqrt{ (\sigma_{published}\times(1+e_2))^2 + (e_1\times flux)^2 }
\label{eq_unccorr}
\end{equation}

The effect of these two parameters $e_1$ and $e_2$ is to increase the uncertainties if they have been underestimated and to add a component to the uncertainty that is proportional to the flux (i.e., a zero-point calibration error). We have adjusted these two parameters manually for each filter band in order to achieve a more Gaussian-like distribution of the photometric residuals. The values are listed in Table \ref{tab_photcorr}. 

\begin{table}
\caption{Adjustments to Photometric Catalog}
  \begin{tabular}{@{}lrrr}
  \hline
Filter & $\Delta$ [mag] & $e_1$ & $e_2$ \\
U&-0.075 &  0.1  &   0.\\
B&+0.00 &  0.05 &   0.\\
V&+0.0125&  0.04 &   0\\
R&+0.025 &  0.03 &   0.\\
I&-0.1 &  0.08 &   0.4\\
z&+0.075 &  0.09 &   -0.4\\
J&-0.15 &  0.15 &   0\\
K&-0.075 &  0.2  &   0.\\
\hline\\
\end{tabular}\\
Parameters for corrections to the published photometric
fluxes and uncertainties, according to Eq. \ref{eq_fluxcorr} and Eq. \ref{eq_unccorr}.
\label{tab_photcorr}
\end{table}

\end{document}